\documentclass[a4paper,11pt]{article}
\pdfoutput=1 

\usepackage{jheppub} 

\usepackage[T1]{fontenc} 

\usepackage{amsmath}
\usepackage{amsfonts}
\usepackage{amssymb}
\usepackage{graphicx}
\usepackage{lmodern}
\usepackage{hyperref}
\usepackage{dcolumn}
\usepackage{bm}
\usepackage[mathlines]{lineno}
\usepackage{float}
\usepackage{subfigure}

\title{\boldmath A unique Neutrino Mass Matrix Texture under Exponential Parametrization}

\author[a]{Pralay Chakraborty,}
\author[a,1]{Subhankar Roy. \note{Corresponding author}}

\affiliation[a]{Department of Physics, Gauhati University, India}

\emailAdd{pralay@gauhati.ac.in}
\emailAdd{subhankar@gauhati.ac.in}

\abstract{Under the exponential parametrization scheme $ m^{\nu}_{ij}\sim r e^{i \theta}$, we propose a \emph{unique} neutrino mass matrix texture that exhibits four correlations among its elements. The mixing scheme obtained from the proposed texture is consistent with experimental observations. We construct a concrete model based on the $SU(2)_L \times U(1)_Y \times A_4 \times Z_{10} \times Z_7$ group within the framework of the seesaw mechanism.}

\begin{document} 
\maketitle
\flushbottom

\newcolumntype{P}[1]{>{\centering\arraybackslash}p{#1}}

\section{Introduction}
\label{introduction}

The neutrino oscillation suggests that neutrinos change their identity as they traverse space \cite{Pontecorvo:1957cp}. This phenomenon can be understood from the quantum mechanics framework which tells us that the three flavour states of neutrinos ($\nu_{l=e,\mu,\tau}$) are the mixture of three mass eigenstates $(\nu_{i=1,2,3})$ having definite masses ($m_{i=1,2,3}$). The neutrino oscillation predicts the existence of non-zero mass, which goes against the predictions of the Standard Model\,(SM) \cite{Glashow:1961tr, Weinberg:1967tq, Salam:1968rm}. In this regard, we need to go beyond the SM framework to understand the non-zero neutrino masses. The neutrino mass matrix ($M_\nu$) originates from the Yukawa Lagrangian, which can be understood from the see-saw mechanism \cite{Yoshimura:1978ex, Akhmedov:1999tm, Schechter:1980gr, Schechter:1981cv, Minkowski:1977sc, Gell-Mann:1979vob, Mohapatra:1979ia, Magg:1980ut, Lazarides:1980nt}. In general, a Majorana neutrino mass matrix contains the information of twelve parameters: three mass eigenvalues ($m_{i=1,2,3}$), three mixing angles ($\theta_{12}, \theta_{13}, \theta_{23}$), three CP phases ($\delta, \alpha, \beta$), and three unphysical phases ($\phi_1, \phi_2, \phi_3$). Recent experiments predict three mixing angles and two mass-squared differences\,($\Delta\,m_{21}^2$ and $\Delta\,m_{31}^2$) precisely \cite{Esteban:2024eli}. However, the octant of $\theta_{23}$, neutrino mass ordering and a precise bound of $\delta$ are yet to be determined. In addition, the oscillation experiments do not measure three individual mass eigenvalues and the Majorana phases. 

However, the $M_\nu$ carries the information of all nine physical parameters, and by imposing certain constraints on $M_\nu$, some of the physical parameters can be predicted. Various constraints, such as texture zeroes \cite{Hagedorn:2005kz, Grimus:2004hf, Xing:2002ta}, vanishing minors \cite{Lashin:2009yd, Dev:2011hf}, hybrid textures \cite{Liu:2013oxa}, and $\mu-\tau$ symmetry \cite{Harrison:2002er} are widely studied in the literature. The experiments rule out the $\mu-\tau$ symmetry due to its prediction of $\theta_{13}$ being zero. In this regard, there are several attempts made to explore deviations from the exact $\mu-\tau$ symmetry \cite{Dey:2022qpu, Chakraborty:2023msb, Chakraborty:2024rgt}. In the present work, we explore different types of constraints, deviating from the existing literature. We adhere to the exponential parametrization of the neutrino mass matrix $M_{\nu_{ij}}=r_{ij} e^{i\theta_{ij}}$, where, $r_{ij}$ represents the absolute value and $\theta$ stands for argument of the matrix elements. In contrast, a complex number $z$ can also be expressed as $z=x+iy$. In principle, physics is independent of parametrization. The underlying principles and phenomena do not alter regardless of the parametrization we choose. However, a suitable parametrization may lead to understand the underlying physics in a better way. Like several other constraints studied in the literature, the said parametrization scheme offers new insights into how physical parameters can be correlated in a different way. The idea of the said parametrization was first discussed in a model-independent way in Ref\,\cite{Chakraborty:2022ess}. The latter discusses several promising textures from different categories, highlighting distinct correlations including the exponential parametrization. All the proposed textures are experimentally viable. However, we have not provided any concrete model in support of the said textures.

In the present work, we put forward a \emph{unique} texture for the neutrino mass matrix highlighting four correlations among the parameters under exponential parametrization, thereby predicting four physical parameters. In addition, following a top-down approach, we derive the texture in its \emph{exact form}, ensuring that the independence of the texture parameters is preserved.

The plan of the paper is outlined as follows: In section (\ref{sec2}), we discuss the details of the proposed texture. In section (\ref{sec3}), we discuss the applications. Section (\ref{sec4}) is devoted to the symmetry realization. In section (\ref{sec5}), we conclude.

\section{Formalism \label{sec2}}

In the present work, we stick to the Majorana nature of neutrino and hence the neutrino mass matrix is expected to be symmetric. In the light of the exponential parametrization, the neutrino mass matrix appears as: $M_{\nu_{ij}}=|m_{ij}|\,e^{i\,\theta_{}ij}$.

Following the said parametrization, we shall try to find out some correlations among the absolute and argument parts of the different matrix elements in a model independent way, which are promising from the phenomenological perspective. 

\subsection{The Proposed Texture}
We posit the following neutrino mass matrix,

\begin{equation}
  M_\nu= \begin{bmatrix}
A & \textbf{r}e^{i \,\boldsymbol{\theta}} & \textbf{r}e^{-i \,\boldsymbol{\theta}}\\
\textbf{r}e^{i \,\boldsymbol{\theta}} & F & G\\
\textbf{r}e^{-i \,\boldsymbol{\theta}} & G & \mathbf{2}\textbf{r}e^{i \,\boldsymbol{\theta}}
\end{bmatrix},
\label{proposed texture}
\end{equation}

where, $A$, $F$ and $G$ are complex parameters. Here, the parameters $A$, $F$ and $G$ are not highlighted in polar form as the correlations among these elements are not sought out. As the above texture showcases four independent correlations, hence we expect four observables to be predicted. 

To draw information on the physical parameters, we need to diagonalize $M: M^d=V^T.M.V$, where, $V$ is a $3\times3$ unitary matrix known as Pontecorvo-Maki-Nakagawa-Sakata (PMNS) matrix \cite{Maki:1962mu}. In general, the PMNS matrix is defined in flavour basis i.e., the basis where the charged lepton mass matrix is diagonal. The Particle Data Group\,(PDG) \cite{ParticleDataGroup:2020ssz} has adopted a parametrization for $V$, where, $V$ is parametrized using three angles and six phases as shown below,

\begin{equation}
\label{pmns}
V= P_{\phi}. U. P_M ,\nonumber
\end{equation}

where, $P_{\phi} =diag (e^{i\phi_1},e^{i\phi_2},e^{i\phi_3})$ and 
$P_M=diag(e^{i\alpha},e^{i\beta},1)$. The matrix $U$ is depicted as shown in the following,

\begin{equation}
U =  \begin{bmatrix}
1 & 0 & 0\\
0 & c_{23} & s_{23}\\
0 & - s_{23} & c_{23}
\end{bmatrix}\times \begin{bmatrix}
c_{13} & 0 & s_{13}\,e^{-i\delta}\\
0 & 1 & 0\\
-s_{13} e^{i\delta} & 0 & c_{13}
\end{bmatrix} \times\begin{bmatrix}
c_{12} & s_{12} & 0\\
-s_{12} & c_{12} & 0\\
0 & 0 & 1
\end{bmatrix},\nonumber
\end{equation}

where, $s_{ij}=\sin\theta_{ij}$ and $c_{ij}=\cos\theta_{ij}$. 

In the present work, we assume that the unphysical phases $\phi_i=1,2,3$ are excluded by redefining the charged lepton fields. In the next section, we shall discuss the experimental viability of the proposed texture.

\subsection{Numerical Analysis}

The four correlations appearing in $M_\nu$ of Eq.\,(\ref{proposed texture}), are manifested as four transcendental equations involving nine physical parameters: $f_i(\theta_{12}, \theta_{13}, \theta_{23}, m_1, m_2, m_3, \delta, \alpha, \beta)=0$, where, $i=1,2.3,4$. We solve these equations by the Newton-Raphson method to generate a sufficient number of random numbers for the parameters $\alpha$, $\beta$, $\delta$ and $\theta_{23}$. For this, we set the mixing angles $\theta_{12}$ and $\theta_{13}$ at their respective $3\sigma$ bound. The mass eigenvalues $m_1$, $m_2$ and $m_3$ are estimated from two mass squared differences $\Delta m_{21}^2$ and $\Delta m_{31}^2$, such that $\Sigma m_i$ is consistent with the cosmological data\,\cite{Planck:2018vyg}. 

\begin{center}
\begin{figure}
\includegraphics[width=1\textwidth]{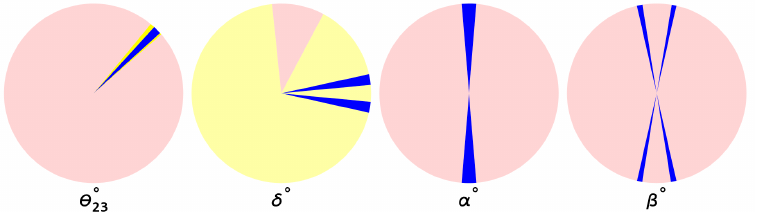}
\caption{The allowed and forbidden regions for the physical parameter based on texture predictions. The blue region signifies the allowed region and yellow region stands for $3\sigma$ region for the studied parameters.}
\label{comparison}
\end{figure}
\end{center}

The analysis leads to some interesting predictions of the observable parameters. We see that $\theta_{23}$ lies within $42^\circ$ to $47^\circ$ approximately (within the $3\sigma$ range\,\cite{Esteban:2024eli}). The $\delta$ experiences two narrow bounds: $5.43^{\circ}$-$12.32^{\circ}$; and  $347.52^{\circ}$-$354.52^{\circ}$. Thus, we see that within the allowed $3\sigma$ bound for $\delta$ given by the experiments, the above texture defines a forbidden region also. Similarly, we see that the Majorana phases $\alpha$ and $\beta$ have got two and four allowed bounds respectively. We see, $\alpha\sim 85.37^\circ - 94.66^\circ$, $265.33^\circ - 274.68^\circ$; and  $\beta\sim 77.32^\circ - 80.33^\circ$, $99.09.33^\circ - 102.78^\circ, 257.35^\circ - 260.74^\circ$, $279.14^\circ - 282.72^\circ$. For a better visualisation of the allowed and forbidden regions of the predicted parameters, we refer to Fig.\,(\ref{comparison}).
 
\begin{figure}
  \centering
    \subfigure[]{\includegraphics[width=0.43\textwidth]{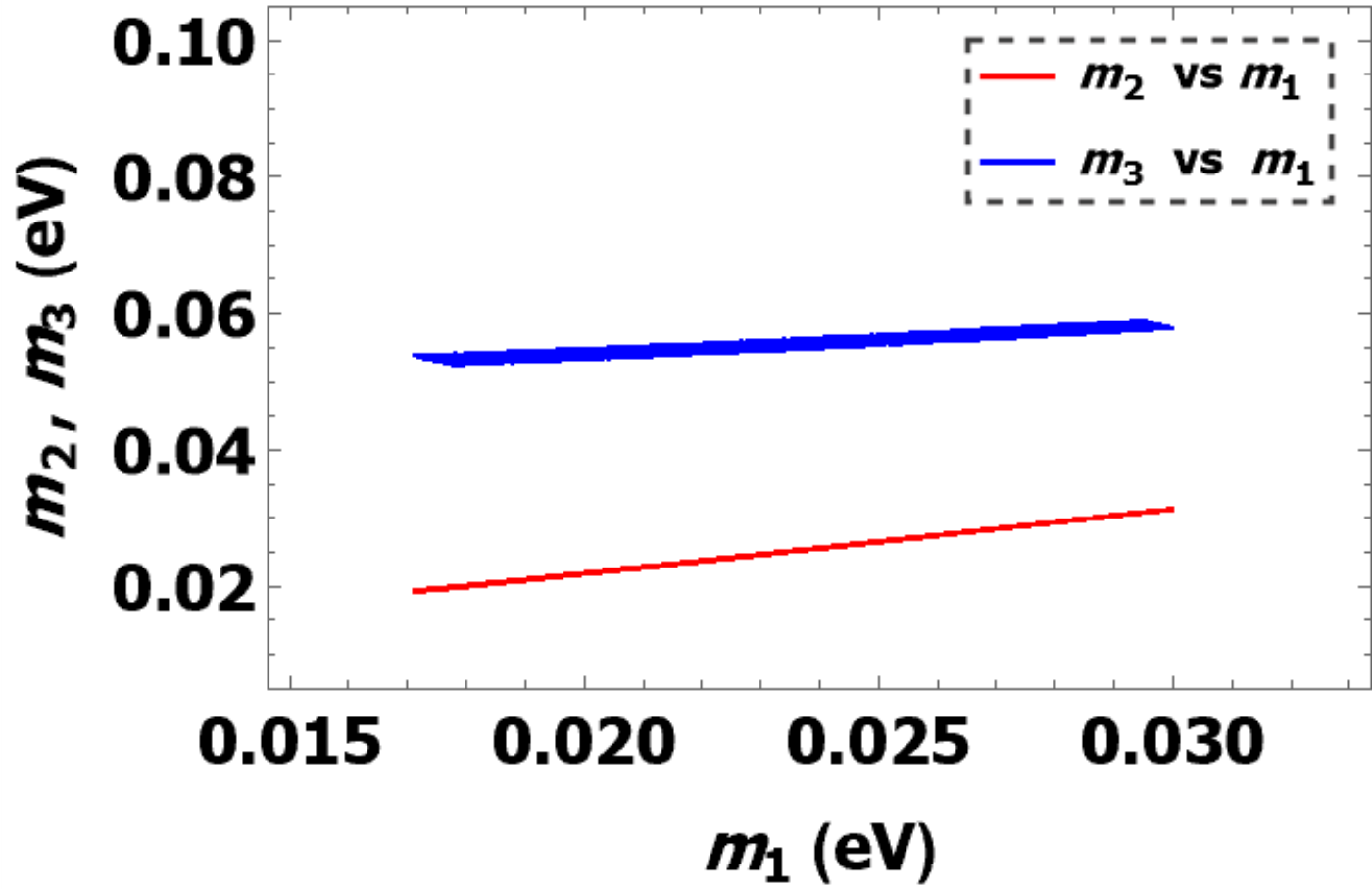}\label{fig:2a}} 
    \subfigure[]{\includegraphics[width=0.43\textwidth]{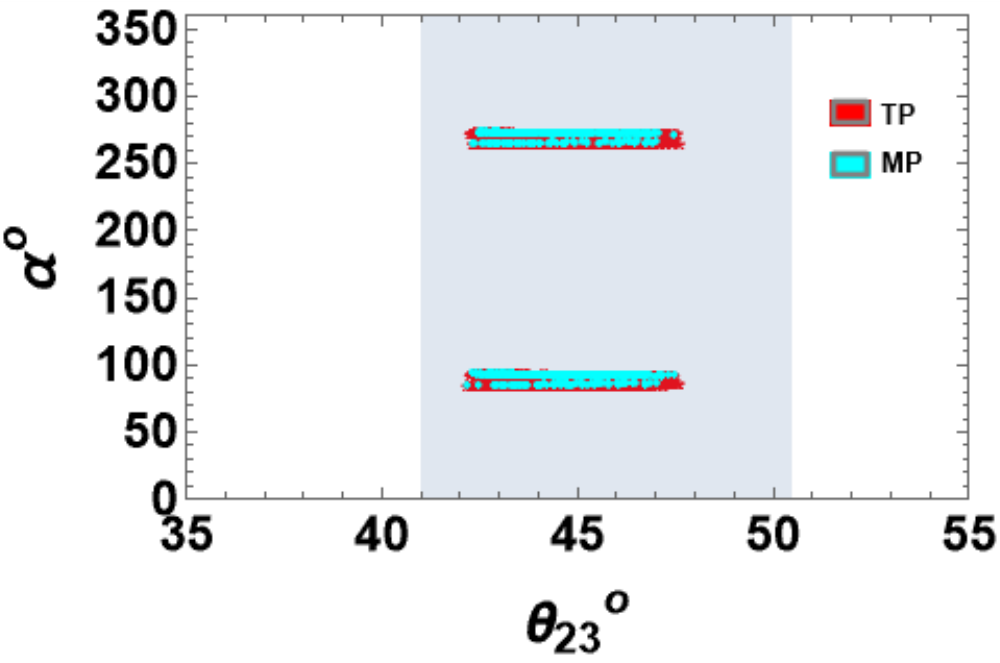}\label{fig:2b}}
    \subfigure[]{\includegraphics[width=0.43\textwidth]{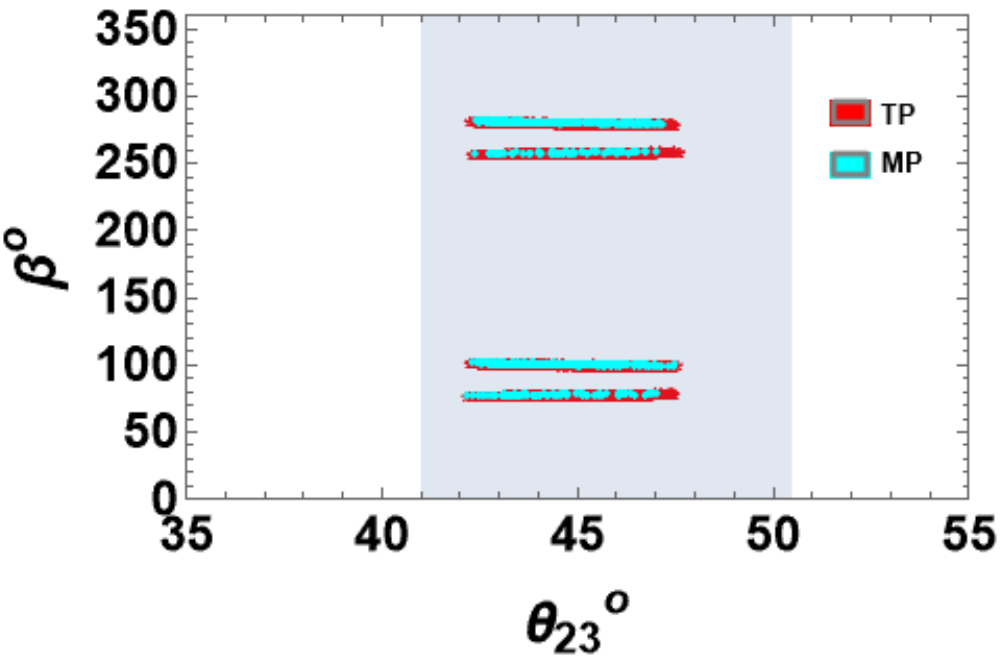}\label{fig:2c}}
    \subfigure[]{\includegraphics[width=0.43\textwidth]{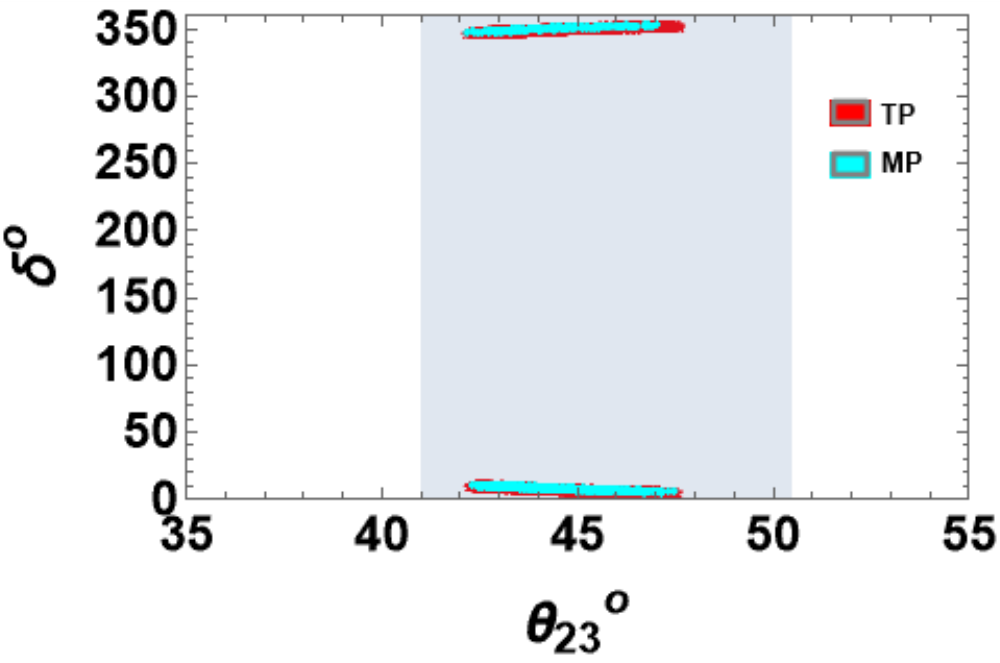}\label{fig:2d}}  
    \subfigure[]{\includegraphics[width=0.45\textwidth]{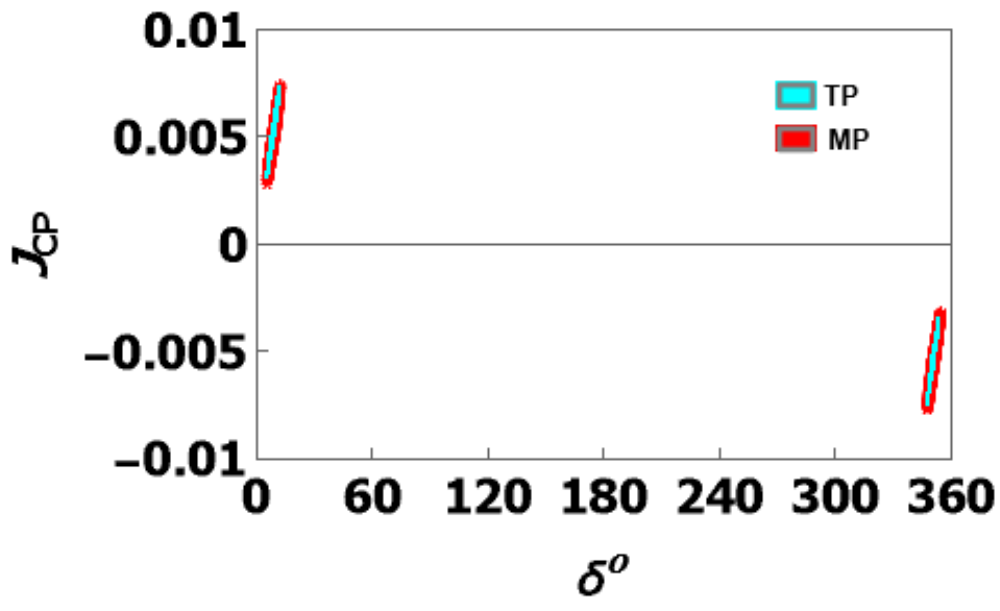}\label{fig:2e}}
    \subfigure[]{\includegraphics[width=0.42\textwidth]{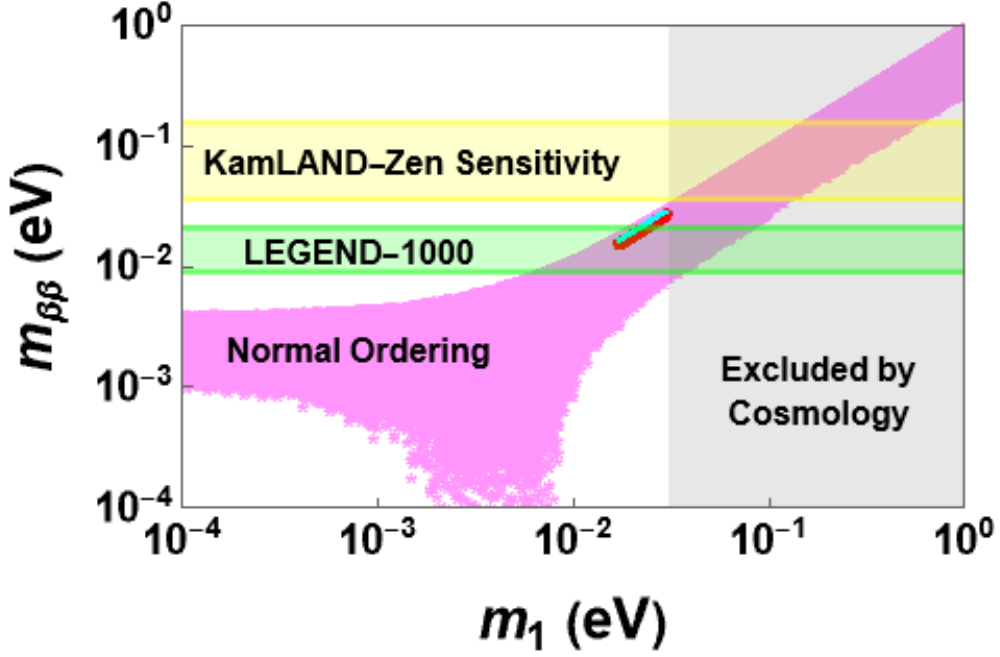}\label{fig:2f}}
   \caption{ Shows the correlation plots between (a) $m_1$ vs $m_2 \,\text{and}\, m_3$. (b) $\alpha$ vs $\theta_{23}$. (c) $\beta$ vs $\theta_{23}$. (d) $\delta$ vs $\theta_{23}$. (e) $J_{CP}$ vs $\delta$. (f) $m1$ vs $m_{\beta\beta}$. The light blue strip represents the $3\sigma$ bound of $\theta_{23}$. The red data points signify texture prediction (TP) while the blue ones signify the corrections from model sector(MP).}
\label{fig:physical parameters}
\end{figure}

Though the mass parameters are taken as input, yet a graphical visualisation of the mass spectra is put forward in Fig.\,(\ref{fig:2a}). In addition, we generate the correlation plots to visualize the predictions of the studied parameters\,(see Figs.\,(\ref{fig:2b})-(\ref{fig:2d})). We know that the Jarlskog invariant ($J_{CP}$) \cite{Jarlskog:1985ht}, that represents the amount of the leptonic CP violation can be obtained in terms of standard parametrization as shown below,
   \begin{equation}
  J_{CP}=s_{13} c_{13}^2 s_{12} c_{12} s_{23} c_{23} \sin\delta \nonumber.
  \end{equation}
  
In this regard, we visualize the variation of $J_{CP}$ against Dirac CP phase $\delta$ from the proposed texture (see Fig.\,\ref{fig:2e}). The maximum and minimum values of the studied parameters are listed in Table\,(\ref{values of physical parameters}). 

It is important to mention that the proposed texture is valid only for the normal ordering of neutrino masses. For inverted ordering, the texture is ruled out as the prediction of $\theta_{23}$ goes outside the experimental bound. 

In the next section, we shall try to analyse the viability of the above texture in the light of other physical parameters.

\begin{table}
\centering
\begin{tabular*}{\textwidth}{@{\extracolsep{\fill}} ccccc}
\hline
\hline
 & Texture  & Texture  & Model  & Model  \\
Parameters & Prediction &  Prediction &Prediction &  Prediction\\
 & (Min Value) & (Max Value) & (Min Value) & (Max Value)\\
\hline
\hline
$m_1 /\text{eV}$ & 0.017 & 0.029 & 0.017 & 0.029\\
\hline
$m_2 /\text{eV}$ & 0.019 & 0.031 & 0.019 & 0.031 \\
\hline
$m_3 /\text{eV}$ & 0.052 & 0.058  & 0.053 & 0.059\\
\hline
$\sum{m_i} /\text{eV}$ &0.09 &0.118  &0.09 &0.119\\
\hline
$\theta_{23}/^\circ$ & 42.17 & 47.61  & 42.39 & 47.48\\
\hline
$\delta/^\circ$ & 5.43 & 354.52  & 5.91 & 354.29\\
\hline
$\alpha/^\circ$ & 85.37 & 274.68   & 85.37 & 274.48\\
\hline
$\beta/^\circ$ & 77.32 & 282.72  & 77.31 & 282.54 \\
\hline
$m_{\beta\beta}/\text{meV}$ & 16.10 & 28.43  & 16.34 & 28.35\\
\hline
\end{tabular*}
\caption{Shows the maximum and minimum values of $m_1$, $m_2$, $m_3$, $\Sigma m_i$, $\theta_{23}$, $\delta$, $\alpha$, $\beta$ and $m_{\beta\beta}$.}
\label{values of physical parameters}
\end{table}

\section{Applications \label{sec3}}

In the earlier section, the proposed texture is found to be consistent with the neutrino oscillation experiments. However, the viability of the texture can be further tested in the light of other physical observables.

\subsection{Effective Majorana Neutrino Mass}

The neutrino-less double beta decay ($0\nu\beta\beta$)\cite{Schechter:1981bd}, is a lepton number violating decay.  Its discovery would justify the Majorana nature of neutrinos. The decay rate $\Gamma$ is determined by the phase space factor $G^{0\nu}$, the nuclear matrix element $M^{0\nu}$, and the effective Majorana mass $m_{\beta\beta}$, following the relation: $\Gamma\sim G^{0\nu}. M^{0\nu}.m^2_{\beta\beta}$.
 
 The effective Majorana neutrino mass $m_{\beta \beta}$ is an observational parameter and it can be expressed as: \[m_{\beta \beta}=|\sum_{k=1}^{3}{U^2_{1k}m_k}|.\] where, $m_1$, $m_2$ and $m_3$ are the three mass eigenvalues. The $U_{11}$, $U_{12}$ and $U_{13}$ are the elements of the PMNS matrix that contains the information of $\alpha$ and $\beta$. Several experiments have provided the upper bounds of $m_{\beta \beta}$: SuperNEMO(Se$^{82}$) as $67-131$ meV,  GERDA(Ge$^{76}$) as $104-228$ meV, EXO-200(Xe$^{136}$) as $111-477$ meV, CUORE(Te$^{130}$) as $75-350$ meV and KamLAND-Zen(Xe$^{136}$) as $61-165$ meV \cite{Ejiri:2020xmm, Agostini:2022zub, CUORE:2019yfd, GERDA:2019ivs, KamLAND-Zen:2016pfg, SuperNEMO:2021hqx, CUORE:2018ncg}. In this regard, we visualize the prediction of the parameter $m_{\beta \beta}$ from the proposed texture\,(see Fig.\,\ref{fig:2f}). It is important to mention that the prediction of $m_{\beta \beta}$ from the proposed texture $(16.10-28.43)$ meV partially falls within the sensitivity of future experiments such as LEGEND-1000 \cite{LEGEND:2021bnm}.

\subsection{CP Asymmetry Parameter}

The effect of leptonic Dirac CP violation $\delta$ on neutrino oscillation can be studied through CP asymmetry parameter $(A_{\mu e})$. The said parameter can be expressed as shown in the following,

\begin{equation}
A_{\mu e}=\frac{P(\nu_\mu \rightarrow \nu_e)-P(\bar{\nu}_\mu \rightarrow \bar{\nu}_e)}{P(\nu_\mu \rightarrow \nu_e)+P(\bar{\nu}_\mu \rightarrow \bar{\nu}_e)}.\nonumber
\end{equation}

The transition probability $P(\nu_\mu \rightarrow \nu_e)$ can be can be expressed in terms of the physical parameters\,\cite{Sinha:2018xof} as shown below,  

\begin{equation}
P(\nu_\mu \rightarrow \nu_e)= P_{\text{atm}}+P_{\text{sol}}+2 \sqrt{P_{\text{atm}}} \sqrt{P_{\text{sol}}} \cos(\Delta_{ij}+\delta),\nonumber
\end{equation}

where, 
$\Delta_{ij}=\Delta m_{ij}^2 \frac{L}{4 E}$. The $E$ represents neutrino beam energy and $L$ stands for baseline length. The quantities $P_{\text{atm}}$ and $P_{\text{sol}}$ can be expressed as shown below,

\begin{eqnarray}
P_{\text{atm}}&=& \sin \theta_{23} \sin 2\theta_{13} \frac{\sin(\Delta_{31}-a L)}{(\Delta_{31}-a L)} \Delta_{31},\nonumber\\
P_{\text{sol}}&=& \cos \theta_{23} \sin 2\theta_{12} \frac{\sin(a L)}{(a L)} \Delta_{21}.\nonumber
\end{eqnarray}

The parameter $a=G_F N_e/\sqrt{2}$ depends on the medium through which the neutrino propagates and arises due to matter effects in neutrino propagation through the Earth. The factor $G_F$ stands for the Fermi constant and $N_e$ represents the electron number density of the medium. For the earth, $a$ is approximately $3500{\rm km}^{-1}$\cite{Sinha:2018xof}. The CP asymmetry parameter can be expressed in terms of the physical parameters as shown below,

\begin{equation}
A_{\mu e}=
\frac{2\sqrt{P_{\text{atm}}}\sqrt{P_{\text{sol}}}\sin\Delta_{32}\sin\delta}{P_{\text{atm}}+2\sqrt{P_{\text{atm}}}\sqrt{P_{\text{sol}}}\cos\Delta_{32}\cos\delta+P_{\text{sol}}}.\nonumber
\end{equation}

\begin{figure}
  \centering
    \subfigure[]{\includegraphics[width=0.33\textwidth]{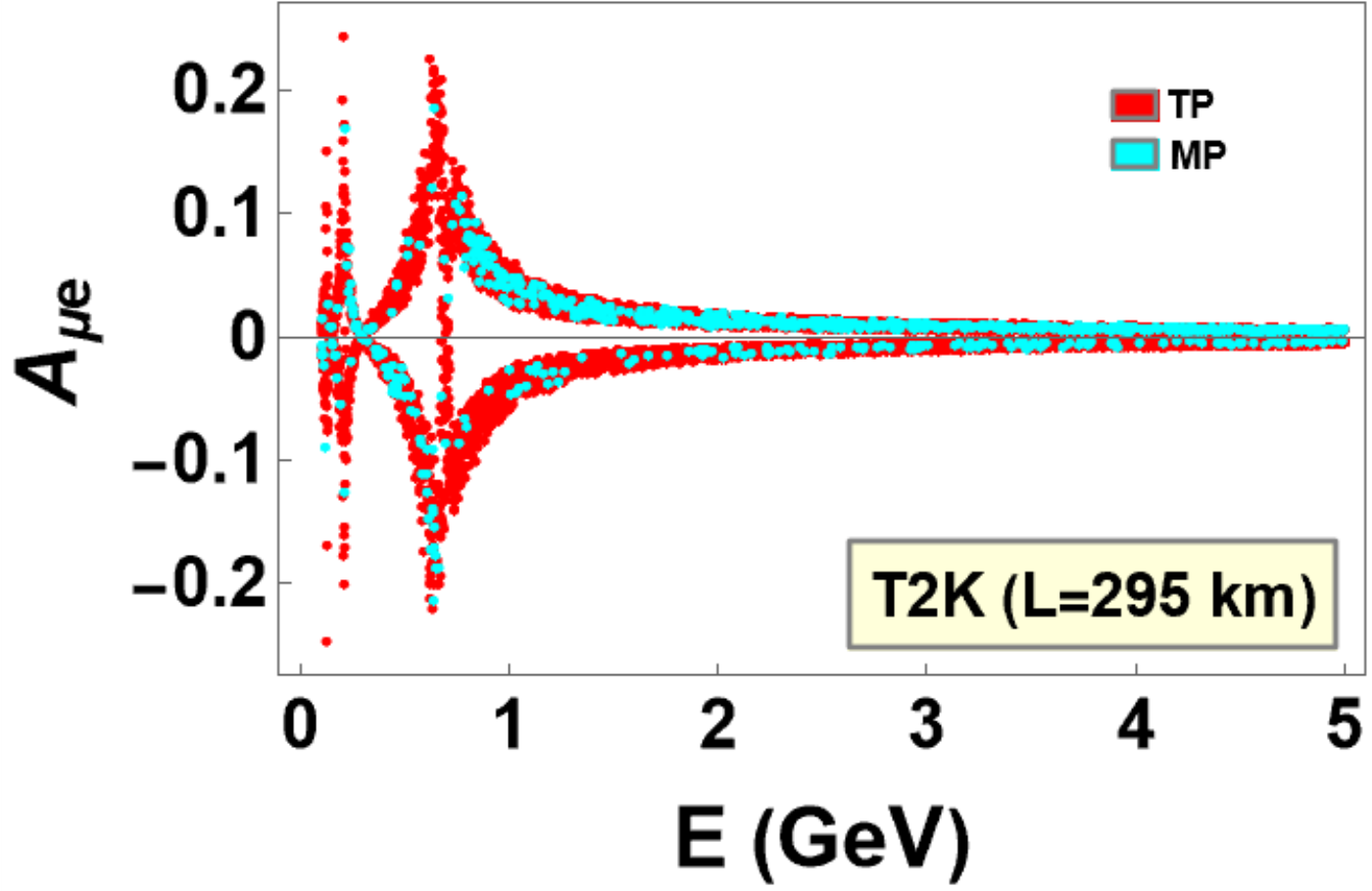}\label{fig:3a}} 
    \subfigure[]{\includegraphics[width=0.32\textwidth]{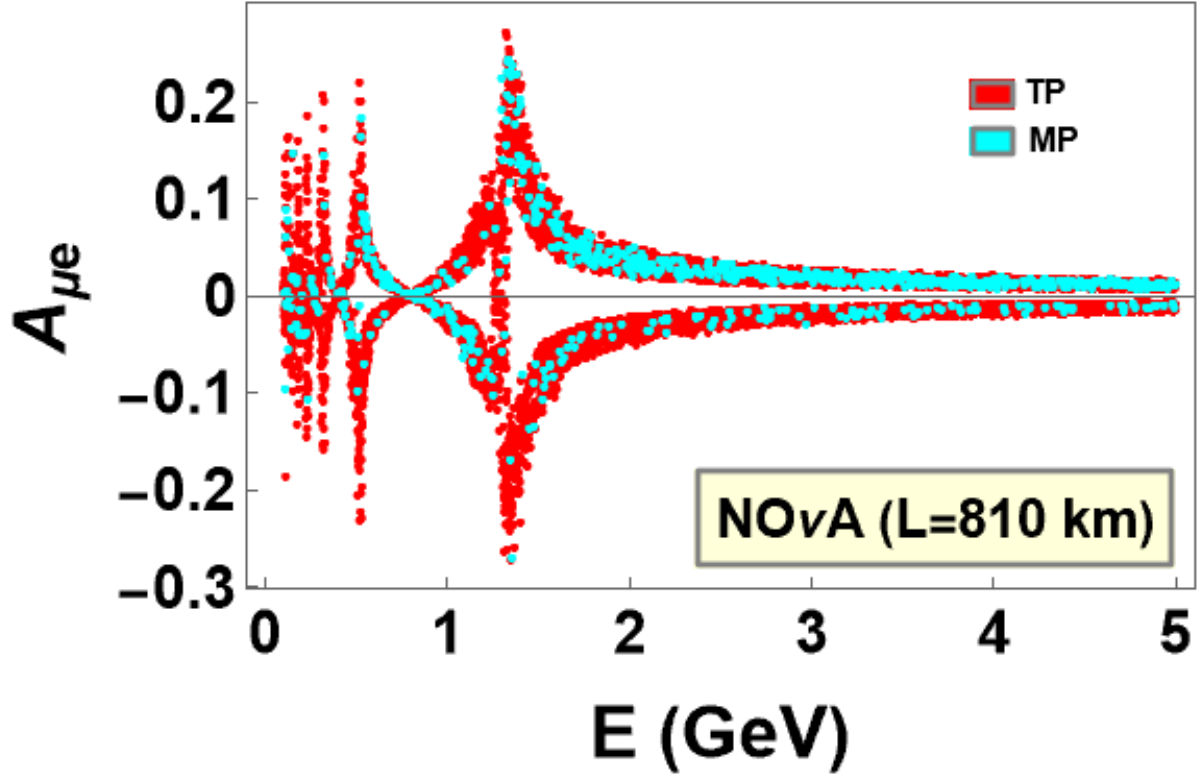}\label{fig:3b}}
    \subfigure[]{\includegraphics[width=0.33\textwidth]{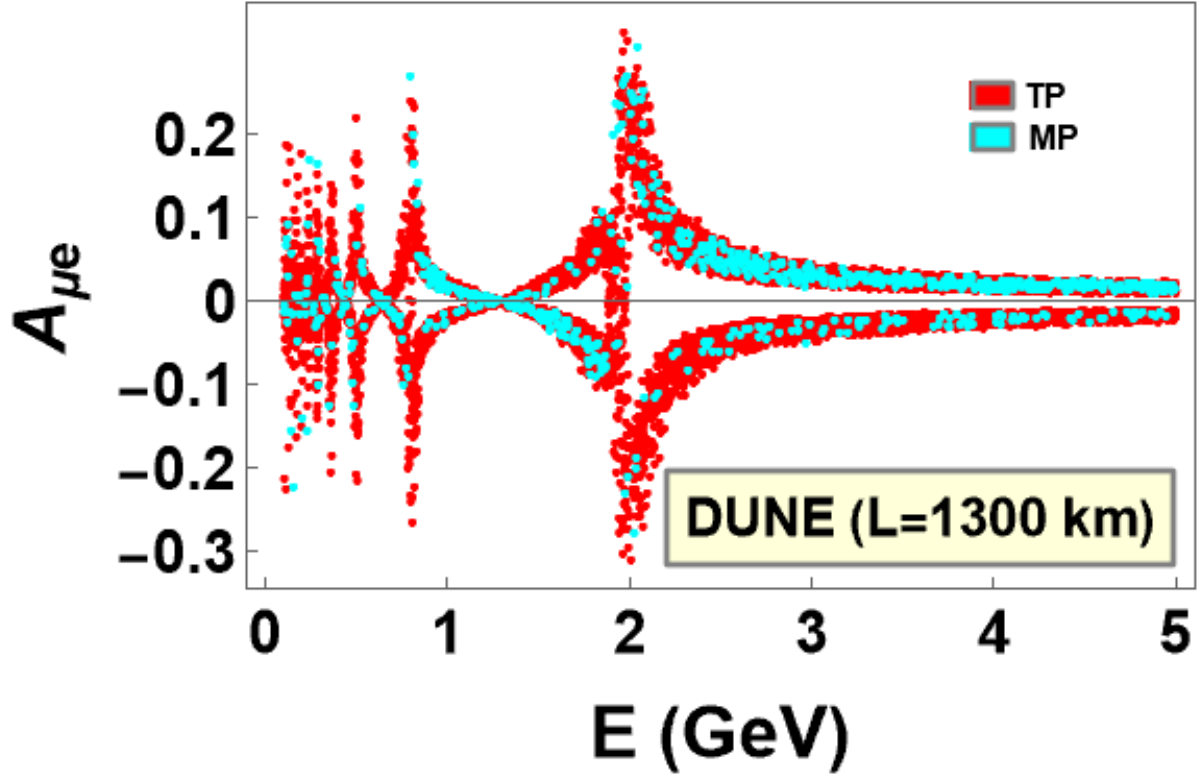}\label{fig:3c}}
   \caption{Shows the correlation plots between (a) $A_{\mu e}$ vs $E$ for T2K $(L=295)$ km. (b) $A_{\mu e}$ vs $E$ for NO$\nu$A $(L=810)$ km. (c) $A_{\mu e}$ vs $E$ for DUNE $(L=1300)$ km. The red data points signify texture prediction (TP) while the blue ones signify the corrections from model sector(MP).}
\label{fig:CP}
\end{figure}

In the present work, we study the CP asymmetry parameter in the light of three experiments with different baseline lengths: T$2$K\,($L=295$ km)\cite{T2K:2019bcf}, No$\nu$A\,($L=810$ km)\cite{NOvA:2021nfi} and DUNE\,($L=1300$ km)\cite{DUNE:2020ypp} (see Figs. \ref{fig:3a}-\ref{fig:3c}). For the above analysis, the beam energy $E$ is set at $(0.5-5)$\,GeV. 

We highlight that the variation of the CP asymmetry parameter is studied from texture predictions, considering different baseline lengths and a range of beam energies. Needless to mention, the parameter $A_{\mu e}$ exhibits a strong dependence on the Dirac CP phase $\delta$. Thus, the long-baseline neutrino experiments are designed to probe this phase, taking into account the matter effects. Notably, the T$2$K and No$\nu$A experiments report differing values for $\delta$ in case of inverted ordering of neutrino masses \cite{NOvA:2021nfi, T2K:2019bcf}. Upcoming experiment like DUNE \cite{DUNE:2020ypp}, with $L=1300$ km and a beam energy peaking around $2.5$ GeV, is expected to provide further insights into this problem, offering a potential test of our prediction for $A_{\mu e}$.

\begin{figure}
  \centering
    \subfigure[]{\includegraphics[width=0.33\textwidth]{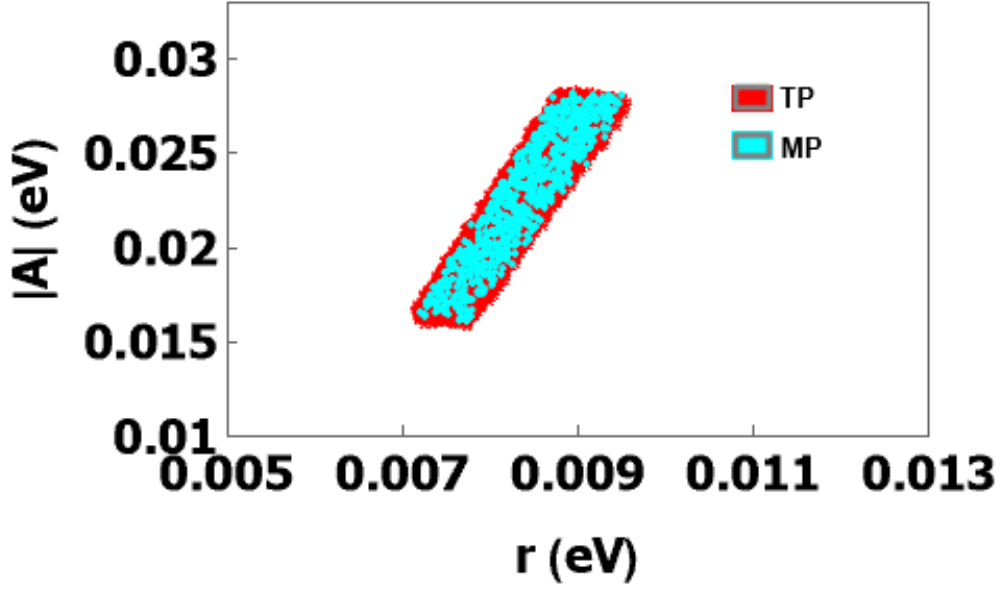}\label{fig:4a}} 
    \subfigure[]{\includegraphics[width=0.33\textwidth]{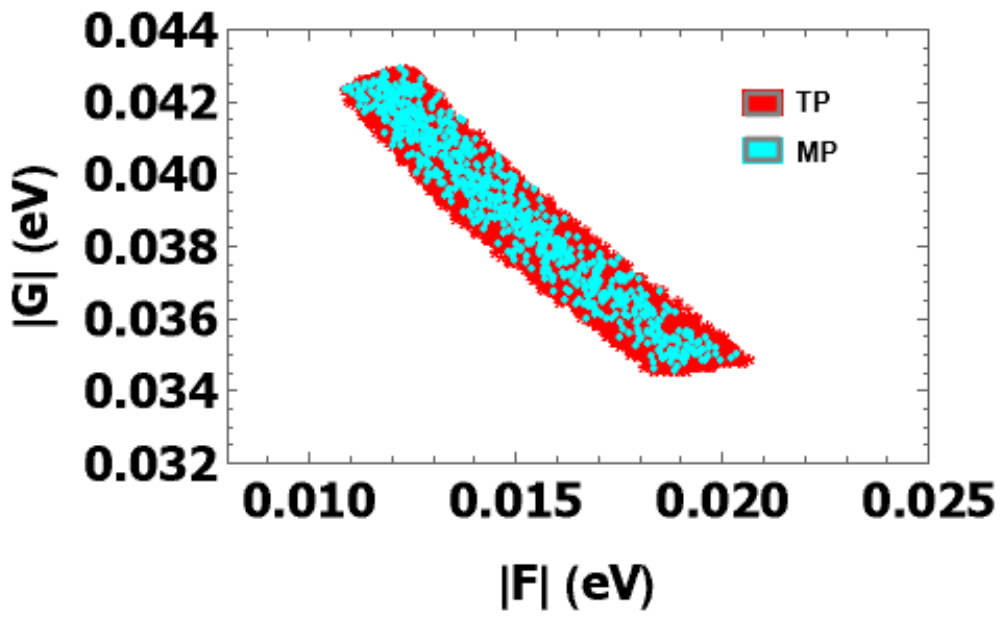}\label{fig:4b}}
    \subfigure[]{\includegraphics[width=0.32\textwidth]{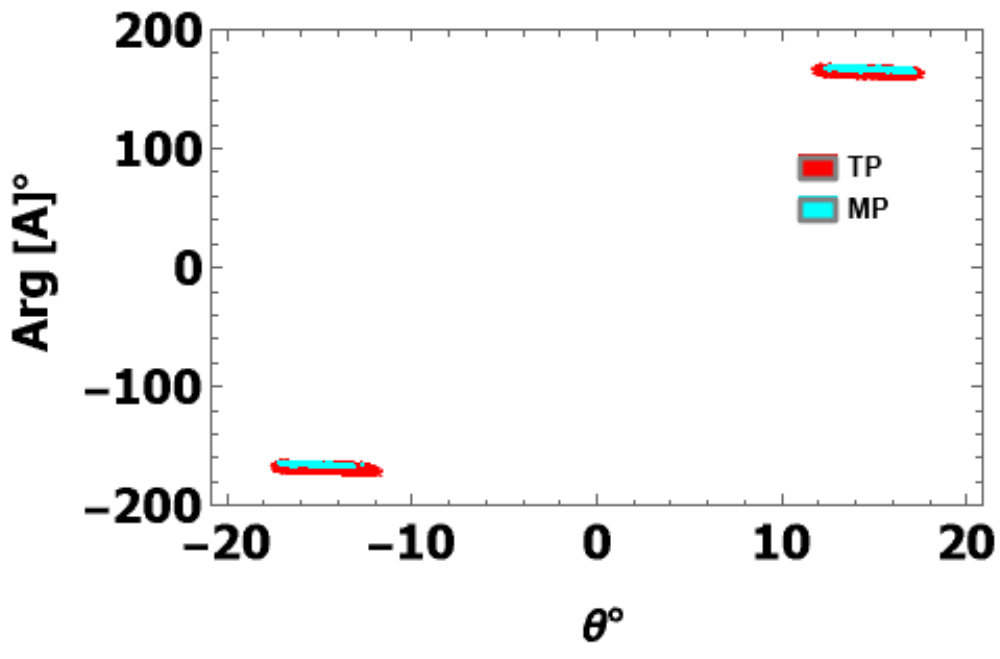}\label{fig:4c}}
    \subfigure[]{\includegraphics[width=0.32\textwidth]{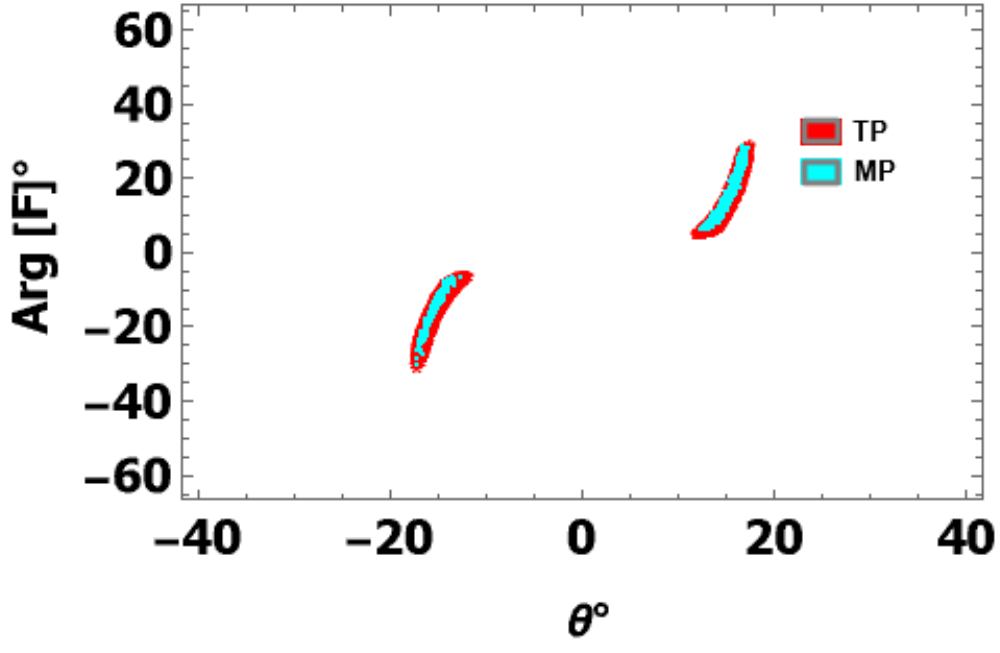}\label{fig:4d}}  
    \subfigure[]{\includegraphics[width=0.32\textwidth]{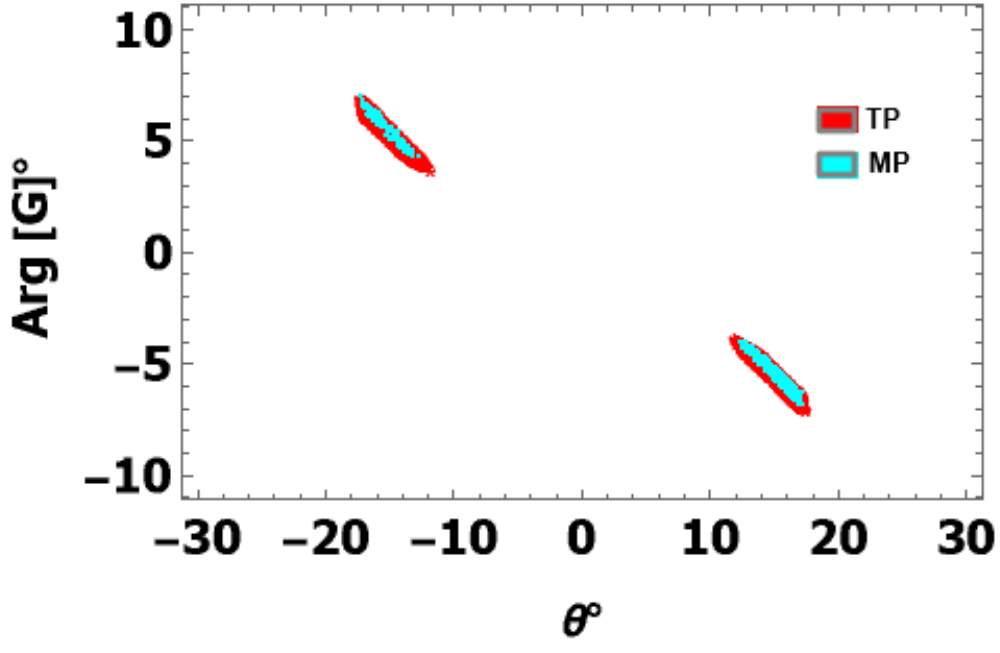}\label{fig:4e}}
   \caption{ Shows the correlation plots between (a) $|A|$ vs $r$. (b) $|G|$ vs $|F|$. (c) $Arg[A]$ vs $\theta$. (d) $Arg[F]$ vs $\theta$. (e) $Arg[G]$ vs $\theta$. The red data points signify texture prediction (TP) while the blue ones signify the corrections from model sector(MP).}
\label{fig:model parameters}
\end{figure}

For the above numerical analysis, the variation of the texture parameters are shown in Fig.\,(\ref{fig:model parameters}). We wish to state that the information on the texture parameters will be helpful while connecting the texture and model sector. In the next section, by following a top-down approach we shall try to achieve the texture from symmetry framework.

\section{Symmetry Framework \label{sec4}}

In section (\ref{sec2}), the texture is proposed in a framework neutral way and the analysis shows that it is consistent with the experimental data. In this section, we present a concrete model for the proposed neutrino mass matrix texture. It is important to note that deriving a model-independent texture within a symmetry framework is quite challenging. Ensuring the independence of the texture parameters is essential while expressing them in terms of the model parameters. To achieve this, we adopt a framework where both Type-I and Type-II seesaw mechanisms contribute. 

\subsection{The Yukawa Lagrangian \label{The Yukawa Lagrangian}}

We start with a a framework where the SM gauge group is extended to $SU(2)_L \times U(1)_Y \times A_4 \times Z_{10} \times Z_7$. In this regard, we extend the field content of the SM by introducing right-handed neutrinos and extra scalar fields. The transformation properties of the extended field content are summarised in Table\,(\ref{Field Content of M}). The product rules under $A_4$ group can be found in \ref{appendix a}. The desired Yukawa Lagrangian borne out of the said framework assumes the following form,

\begin{eqnarray}
- \mathcal{L}_Y &=& y_{e}(\bar{D}_{l_{L}}H) e_{R} + y_{\mu}(\bar{D}_{l_{L}}H)\mu_{R} + y_{\tau}(\bar{D}_{l_{L}}H)\tau_{R}+ \,y_{1}(\bar{D}_{l_{L}}\tilde{H})\nu_{eR}+\frac{y_2}{\Lambda}\nonumber\\&&(\bar{D}_{l_{L}}\tilde{H})\nu_{\mu_R} \chi + \,\frac{y_3}{\Lambda} (\bar{D}_{l_{L}}\tilde{H})\nu_{\tau_R}\psi+ \frac{y_{a}}{2}(\overline{\nu^c}_{e_{R}}\nu_{e_{R}})\xi+ \frac{y_{b}}{2\Lambda}[(\overline{\nu^c}_{\mu_{R}}\nu_{\tau_{R}})\nonumber\\&& + (\overline{\nu^c}_{\tau_{R}}\nu_{\mu_{R}})]\xi\kappa+ \frac{y_{c}}{2}(\overline{\nu^c}_{\mu_{R}}\nu_{\mu_{R}})\eta + y_{T_{2}}(\bar{D}_{l_{L}} D_{l_{L}}^{c})\,i \sigma^2 \Delta + h.c.,
\label{Yukawa Lagrangian M1}
\end{eqnarray}

where, $\Lambda$ is the cut-off scale of the theory. The Lagrangian is constructed in leading order. The presence of $Z_{10}$ \cite{Dey:2023bfa} in the model restricts the undesirable terms that are permitted by $A_4$. Keeping in mind the targeted neutrino mass matrix in Eq.\,(\ref{proposed texture}), the vacuum expectation values\,(vev) for the scalar fields are chosen as: $\langle H \rangle_{0}=v_{H}(1,1,1)^{T}$, $\langle \Delta \rangle_{0}=v_{\Delta}(v_1,v_2,0)^{T}$, $\langle\chi\rangle_{0}=v_{\chi}$, $\langle\psi\rangle_{0}=v_{\psi}$, $\langle\xi\rangle_{0}=v_{\xi}$, $\langle\eta\rangle_{0}=v_{\eta}$ and $\langle\kappa\rangle_{0}=v_{\kappa}$ respectively. The potential that shelters the chosen vacuum alignments is discussed in subsection \ref{secalar potetial}.

\begin{table}[H]
\centering
\begin{tabular}{P{1.5cm}P{0.5cm}P{2cm}P{2cm}P{0.7cm}P{0.7cm}P{0.7cm}P{0.7cm}P{0.7cm}P{0.7cm}P{0.8cm}} 
\hline
Fields & $D_{l_{L}}$ & $l_{R}$ & $\nu_{l_R}$ & $H$ & $\Delta$ & $\chi$ & $\psi$ & $\xi$ & $\eta$ &  $\kappa$ \\ 
\hline\hline
$SU(2)_{L}$ & 2 & 1 & 1 & 2 & 3 & 1 & 1 & 1 & 1 & 1 \\
\hline
$U(1)_Y$ & -1 & $(-2,-2,-2)$ & $(0,0,0)$ & $1$ & -2& 0 & 0 & 0 & 0 & 0  \\
\hline
$A_4$ & 3 & $(1,1',1'')$ & $(1,1',1^{''})$ & 3 & 3 & 1 & 1 & 1 & $1'$ & 1  \\
\hline
$Z_{10}$ & 0 & $(0,0,0)$ & $(0,4,3)$ & $0$ & 0 & 6 & $7$ & $0$ & $2$ & 3  \\
\hline
$Z_{7}$ & -2 & $(4,4,4)$ & $(6,6,6)$ & 1 & 3 & 0 & 0 & 2 & 2 & 0  \\
\hline
\end{tabular}
\caption{The transformation properties of various fields under $SU(2)_L \times U(1)_Y \times A_4 \times Z_{10} \times Z_7$ group.} 
\label{Field Content of M}
\end{table}

 We derive the charged lepton mass matrix($M_{l}$) as shown in the following,

	\begin{equation}
		M_{l}=v_{H} \begin{bmatrix}
			y_{e} & y_{\mu} & y_{\tau} \\
			y_{e} & \omega\,y_{\mu} & \omega^2\, y_{\tau} \\
			y_{e} & \omega^2\, y_{\mu} & \omega \,y_{\tau} \\
		\end{bmatrix},\nonumber
	\end{equation}

	We diagonalize $M_l$ as $M_{l}^{diag}=U_{l_{L}}^{\dagger}M_{l}U_{l_R}$, where $M^{diag}_{l}=\sqrt{3}\,v_H \,diag\,(y_{e},y_{\mu},y_{\tau})$. The $U_{l_{L}}$ and $U_{l_{R}}$ are expressed as shown below,
	\begin{equation}
		U_{l_{L}}=\frac{1}{\sqrt{3}} \begin{bmatrix}
			e^{i\rho} &  1 & 1\\
			e^{i\rho} & \omega & \omega^2\\
			e^{i\rho} & \omega^2 & \omega\\
		\end{bmatrix},\quad
		U_{l_{R}}=diag\begin{bmatrix}
			e^{i\rho}, & 1, & 1
		\end{bmatrix}.\nonumber
	\end{equation}

Here, we include a phase $\rho$ in $U_{l_{L}}$ and $U_{l_{R}}$, motivated by the fact that the choices of the eigenvectors are not unique \cite{Dey:2022qpu, Chakraborty:2023msb,Chakraborty:2024rgt, Goswami:2025jde}. The presence of this arbitrary phase does not alter the original structure of $M_l$ rather specifies a preferred charged lepton basis. The presence of the phase in $U_{l_{L}}$ will certainly affect the neutrino mass matrix in flavour basis, i.e., the basis where the $M_l$ is diagonal. Thus, a proper choice of the phase will help us to obtain the desired neutrino mass matrix. In the present work, we set $\rho=\pi$.

The Dirac neutrino mass matrix($M_{D}$) and the right-handed neutrino mass matrix($M_{R}$) are obtained as follows,
	
	\begin{equation}
		M_{D}= \begin{bmatrix}
			y_1 v_{H} & \frac{y_2}{\Lambda} v_{H} v_\chi &  \frac{y_3}{\Lambda} v_{H} v_\psi \\
			y_1 v_{H} & \frac{y_2}{\Lambda} \omega v_{H} v_\chi & \frac{y_3}{\Lambda} \omega^{2} v_{H} v_\psi \\
			y_1 v_{H} & \frac{y_2}{\Lambda} \omega^{2} v_{H} v_\chi & \frac{y_3}{\Lambda} \omega v_{H} v_\psi \\
		\end{bmatrix},\quad
	M_{R}= \begin{bmatrix}
			\frac{y_a}{2} v_\xi & 0 & 0 \\
			0 & \frac{y_c}{2} v_\eta & \frac{y_b}{2\Lambda} v_\xi v_\kappa \\
			0 & \frac{y_b}{2\Lambda} v_\xi v_\kappa & 0\\
		\end{bmatrix}.\nonumber
	\end{equation}
	
We take the Type-I and Type-II contributions and construct the light neutrino mass matrix ($M_{\nu s}$) in the basis where $M_l$ is non-diagonal
  
	\begin{equation}
		M_{\nu s}= -M_{D}M_{R}^{-1}M_{D}^{T}\,\,\,+\,\, \begin{bmatrix}
			0 & 0 & y_{T_{2}}v_2\\
			0 & 0 & y_{T_{2}}v_1 \\
			y_{T_{2}}v_2 & y_{T_{2}} v_1  & 0\\
		\end{bmatrix}.\nonumber
	\end{equation}

On redefining the above neutrino mass matrix in a basis where $M_l$ is diagonal, we obtain the final neutrino mass matrix following the transformation: $M_{\nu}=U_{l_{L}}^{T}M_{\nu_{s}}U_{l_{L}}$ as shown below, 

\begin{equation}
\scriptstyle
  M_{\nu}= \begin{bmatrix}
-\frac{6 v^2 y_1^2}{y_a v_{\xi }}+\frac{2 v_1 y_T}{3}+\frac{2 v_2 y_T}{3} & \frac{ y_T}{3} \sqrt{v_1^2-v_2 v_1+v_2^2}\,e^{i \phi} & \frac{ y_T}{3} \sqrt{v_1^2-v_2 v_1+v_2^2}\,e^{-i \phi}\\
 \frac{ y_T}{3} \sqrt{v_1^2-v_2 v_1+v_2^2}\,e^{i \phi} & \frac{6 v^2 y_3^2 y_c v_{\eta } v_{\psi }^2}{\Lambda ^2 y_b^2 v_{\kappa }^2 v_{\xi }^2}+y_T(\frac{2 v_1 }{3}-\frac{v_2}{3}-\frac{i v_2}{\sqrt{3}}) & -\frac{6 v^2 y_2 y_3 v_{\chi } v_{\psi }}{\Lambda ^2 y_b v_{\kappa } v_{\xi }}-\frac{ y_T}{3}(v_1+v_2)\\
\frac{ y_T}{3} \sqrt{v_1^2-v_2 v_1+v_2^2} \,e^{-i \phi} & -\frac{6 v^2 y_2 y_3 v_{\chi } v_{\psi }}{\Lambda ^2 y_b v_{\kappa } v_{\xi }}-\frac{ y_T}{3}(v_1+v_2) & \frac{2 y_T}{3} \sqrt{v_1^2-v_2 v_1+v_2^2}\, e^{i \phi}
\end{bmatrix},
\label{model texture}
\end{equation} 	

where, $\phi=-\tan^{-1}\left[\sqrt{3} v_2/(v_2-2 v_1)\right]$. The above model enhanced mass matrix carrying \emph{sixteen} model parameters involving eight vevs, seven Yukawa couplings and the cut-off scale, resembles the exact form of the neutrino mass matrix texture expressed in terms of $r, \theta, A, F$ and $G$ as presented in Eq.\,(\ref{proposed texture}). It's worth mentioning that the independence of the texture parameters are maintained while expressing them in terms of the model parameters. 

In principle, the Weinberg like operator,

\begin{equation}
\mathcal{L}_W = \frac{1}{\Lambda} \bar{D}_{l_{L}} \tilde{H} \tilde{H}^T D_{l_{L}}^{c}\nonumber,
\end{equation}

may appear in the model\,\cite{Vien:2025fiu}. In order to restrict the said terms in our model, the $Z_7$ symmetry is invoked. 

We wish to highlight that the present symmetry framework is designed in order to obtain the proposed texture of neutrino mass matrix in Eq.\,(\ref{proposed texture}) in its exact form. To achieve this, both type-I and type-II seesaw mechanisms are incorporated. However, this enhances the field content. The texture parameters including $r$, $\theta$ and the other complex parameters $A$, $F$ and $G$ are all independent. The model parameters involving the Yukawa couplings and vevs etc. combine in specific ways  and resemble the texture parameters. However, this step is challenging as this transformation might lead to the loss of independence of the texture parameters. To restrict this, the entry of additional scalar fields is inevitable. However, we don't deny that another minimalistic approach to the problem is  possible.

In the next section, we shall try to relate the parameters from texture and model sector.

\subsection{Important Relations}	 
	
\begin{figure}
  \centering
    \subfigure[]{\includegraphics[width=0.34\textwidth]{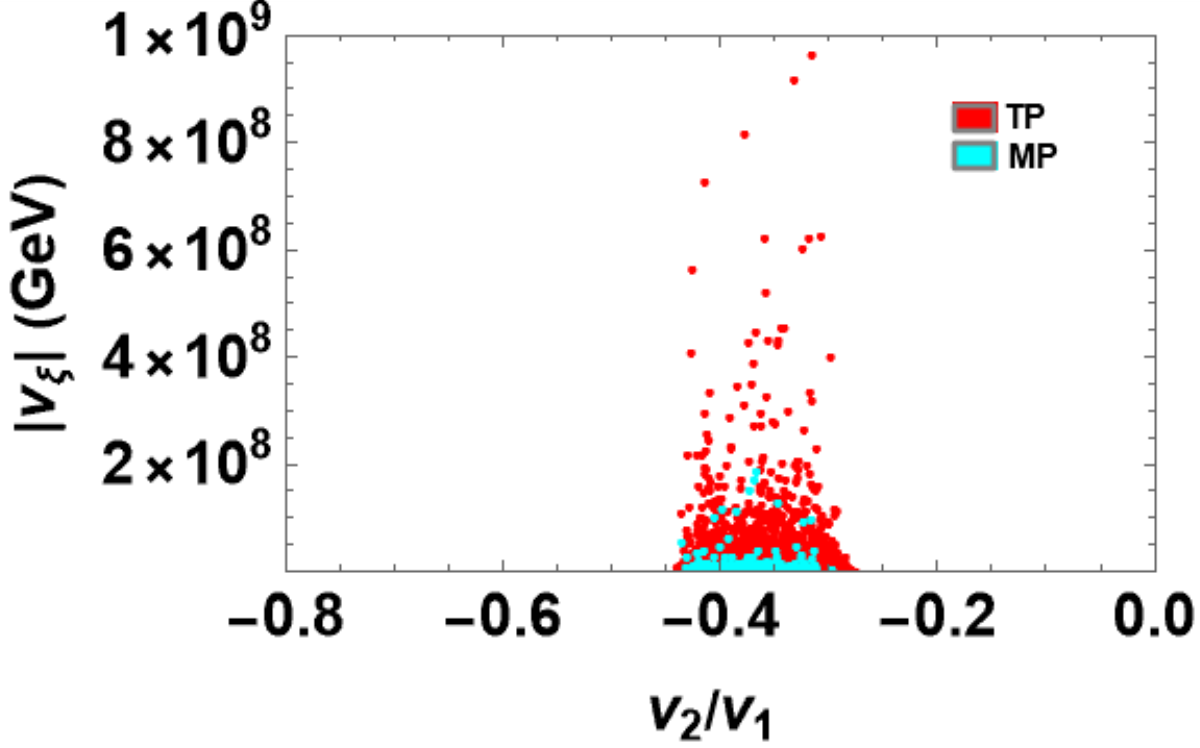}\label{fig:5a}} 
    \subfigure[]{\includegraphics[width=0.31\textwidth]{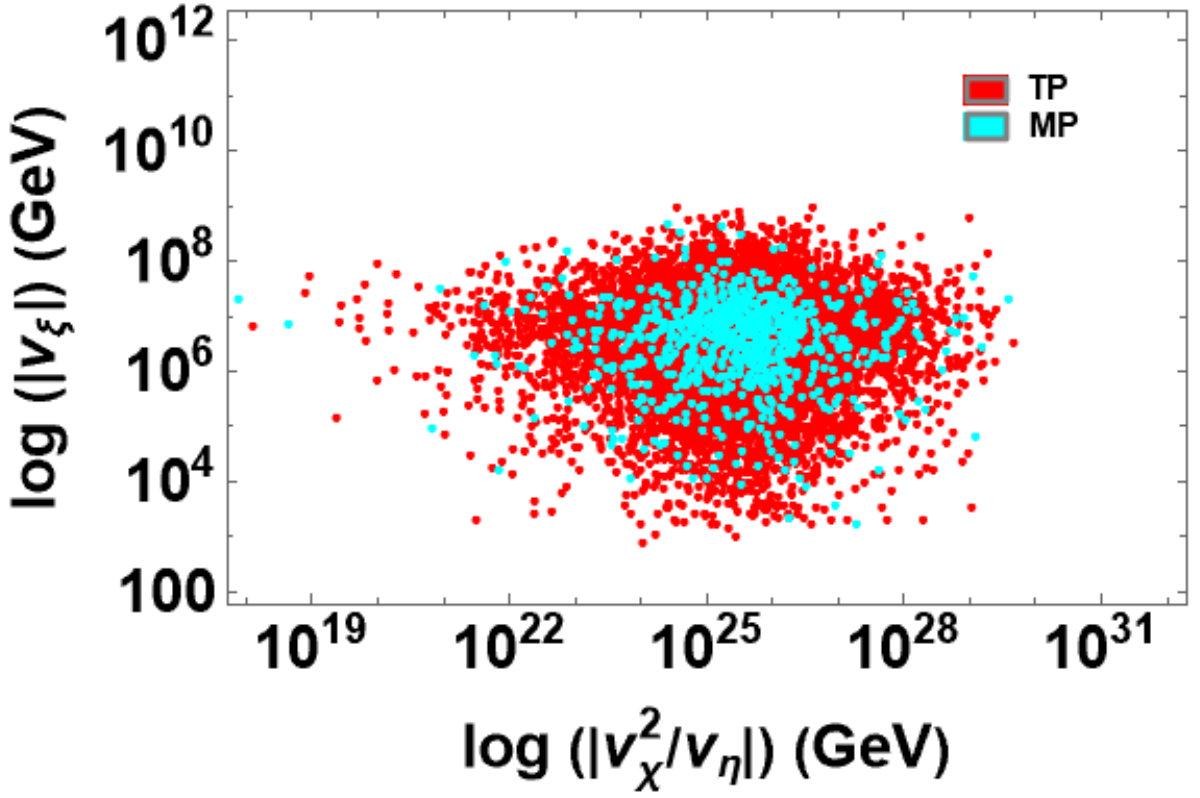}\label{fig:5b}}
    \subfigure[]{\includegraphics[width=0.31\textwidth]{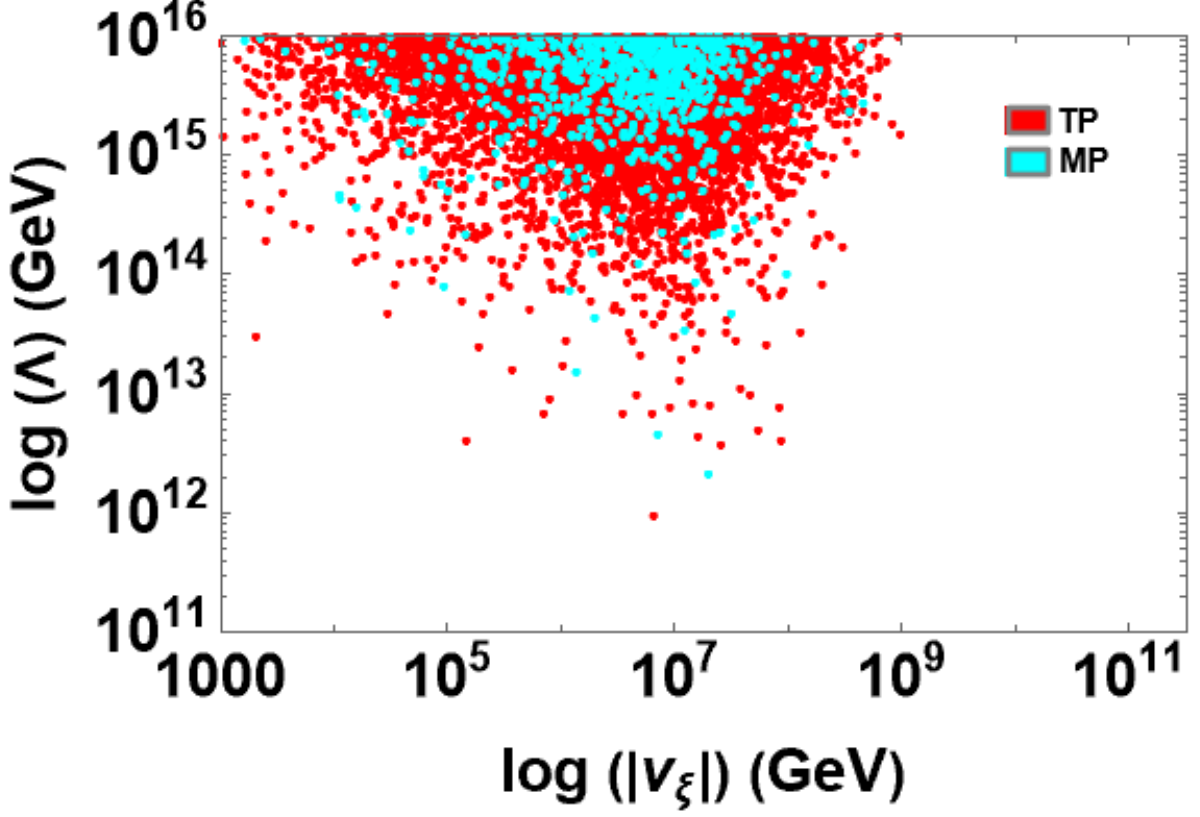}\label{fig:5c}}
   \caption{ Shows the correlation plots between (a) $v_1/v_2$ vs $|v_\xi|$. (b) $\log(|v_\xi|)$ vs $\log(|v_\chi^2/v_\eta|)$. (c) $\log(\Lambda)$ vs $\log(|v_\xi|)$. The red points signify texture prediction (TP) while the blue data points signify the corrections from model sector(MP).}
\label{fig:ratio}
\end{figure}

We highlight that, in order to obtain the proposed texture as shown in Eq.\,(\ref{proposed texture}) from symmetry, the $v_1$ and $v_2$ must be real. In addition, the desired texture demands Yukawa coupling $y_T$ as positive. For integrity, the numerical values of the individual model parameters are needed to be recognised. However, this task is challenging. However, we obtain some useful relations among the model and texture parameters as shown below,
	
	\begin{eqnarray}
	\frac{v_1}{v_2}&=& 2-\frac{6 b_1}{3 b_1+\sqrt{3}b_2},\label{c1}\\
	\frac{v_\chi^2}{v_\eta}&=&\frac{\Lambda^2  y_c \left(b_1+\sqrt{3} b_2+g\right)^2}{6 v^2 y_2^2 (-2 b_1+2 i b_2+f)},\\
	v_\xi &=& -\frac{6 \sqrt{3} v^2 y_1^2}{\sqrt{3} a y_a-2 \sqrt{3} b_1 y_a-6 b_2 y_a},\label{c3}
	\end{eqnarray}
	
where, $b_1=r \cos \theta$ and $b_2=r \sin \theta$. We expect that these relations, somehow, will help to understand the numerical domain of the model parameters more closely. The parameter space of the elements, $v_1/v_2$, $|v_\chi^2/v_\eta|$ and $|v_\xi|$ are presented in Fig.\,\ref{fig:ratio}). For this, we allow the cut-off scale and the Yukawa couplings to vary randomly in the bounds: $(10^3-10^{16})$ GeV and $(0.01-0.99)$ respectively. Needless to mention, the $v_H$ is taken as $246$ GeV.

We mention that the relations appearing in Eqs.\,(\ref{c1})-(\ref{c3}) are solely obtained from the Yukawa sector. However, the vevs that appear there, are solely connected to the scalar sector, which will be addressed in the next section.

\subsection{The Scalar Sector \label{secalar potetial}}

There are seven $A_4$ complex scalar field multiplets in the model and, the associated  $SU(2)_L \times U(1)_Y \times A_4 \times Z_{10} \times Z_7$ invariant potential, is shown explicitly in \ref{appendix b}. We see that there are seven $\mu^2$'s, forty $\lambda$'s and one $\mu$ in the potential expression. Thus it enhances the liberty to choose the desired vacuum alignments for the scalar fields. The eleven minimisation conditions in support of the vacuum alignments as discussed in section\,\ref{The Yukawa Lagrangian} are elaborated in \ref{appendix b}.

From the minimization conditions, we obtain the following significant relations:

\begin{eqnarray}
\lambda_4^{H \Delta}&=& \frac{3}{4} \lambda_2^{H \Delta} \left(\frac{1}{(\frac{v_1}{v_2})^2}+\left(\frac{v_1}{v_2}\right)-\frac{2}{(\frac{v_1}{v_2})}+1\right)\label{potential condition 1},\\
\mu v_2 &=& \frac{2 \lambda_3^{H \Delta} \left(\left(\frac{v_1}{v_2}\right)^2-\left(\frac{v_1}{v_2}\right)+1\right)}{\left(\frac{v_1}{v_2}\right)^2-3 \left(\frac{v_1}{v_2}\right)-\frac{1}{\left(\frac{v_1}{v_2}\right)}+2}\label{potential condition 2},\\
\left(\frac{v}{v_2}\right)^2 &=& \frac{\left(\frac{v_1}{v_2}\right)^3 \left(\left(\left(\frac{v_1}{v_2}\right)^2 - 1\right) (3 \lambda_{2}^\Delta - 4 \lambda_{3}^\Delta) - 2 \lambda_{4}^\Delta \left(\left(\frac{v_1}{v_2}\right)^2 + 1\right)\right)}{3 \left(\lambda_{2}^{H \Delta} + \lambda_{2}^{H \Delta} \left(\frac{v_1}{v_2}\right) \left(\left(\frac{v_1}{v_2}\right) \left(\left(\frac{v_1}{v_2}\right)^3 - 2 \left(\frac{v_1}{v_2}\right) + 4\right) - 3\right)\right)}
\label{potential condition 3}.
\end{eqnarray}

The above relations are significant in the sense that those are expressible in terms of $v_1/v_2$, which, in turn, is dictated by the texture parameters $r$ and $\theta$(see Eq.\,(\ref{c1})). In the next section, we shall see how these relations will help us to curtail the parameter space.

\begin{figure}[H]
  \centering
    \subfigure[]{\includegraphics[width=0.4\textwidth]{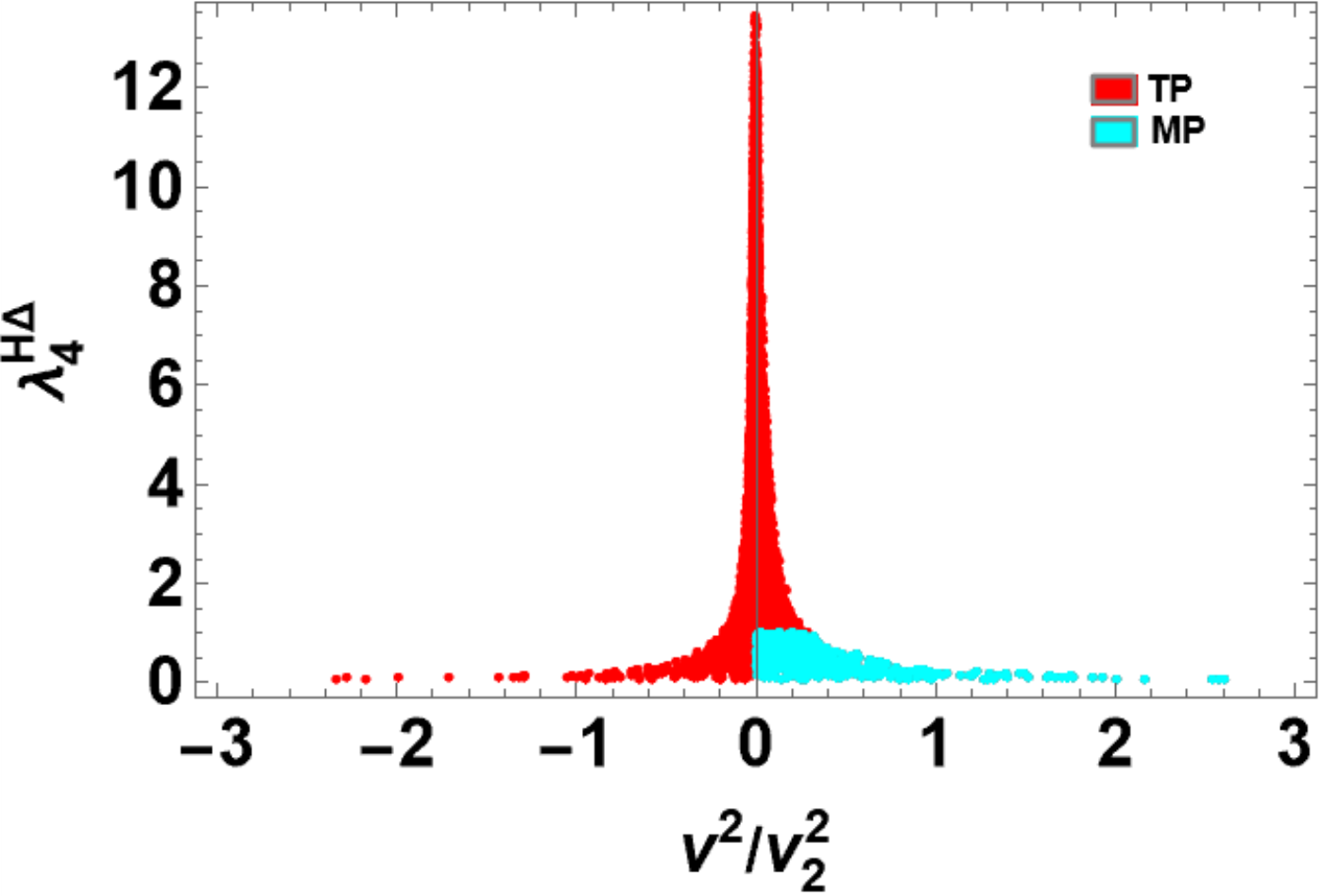}\label{fig:6a}} 
    \subfigure[]{\includegraphics[width=0.41\textwidth]{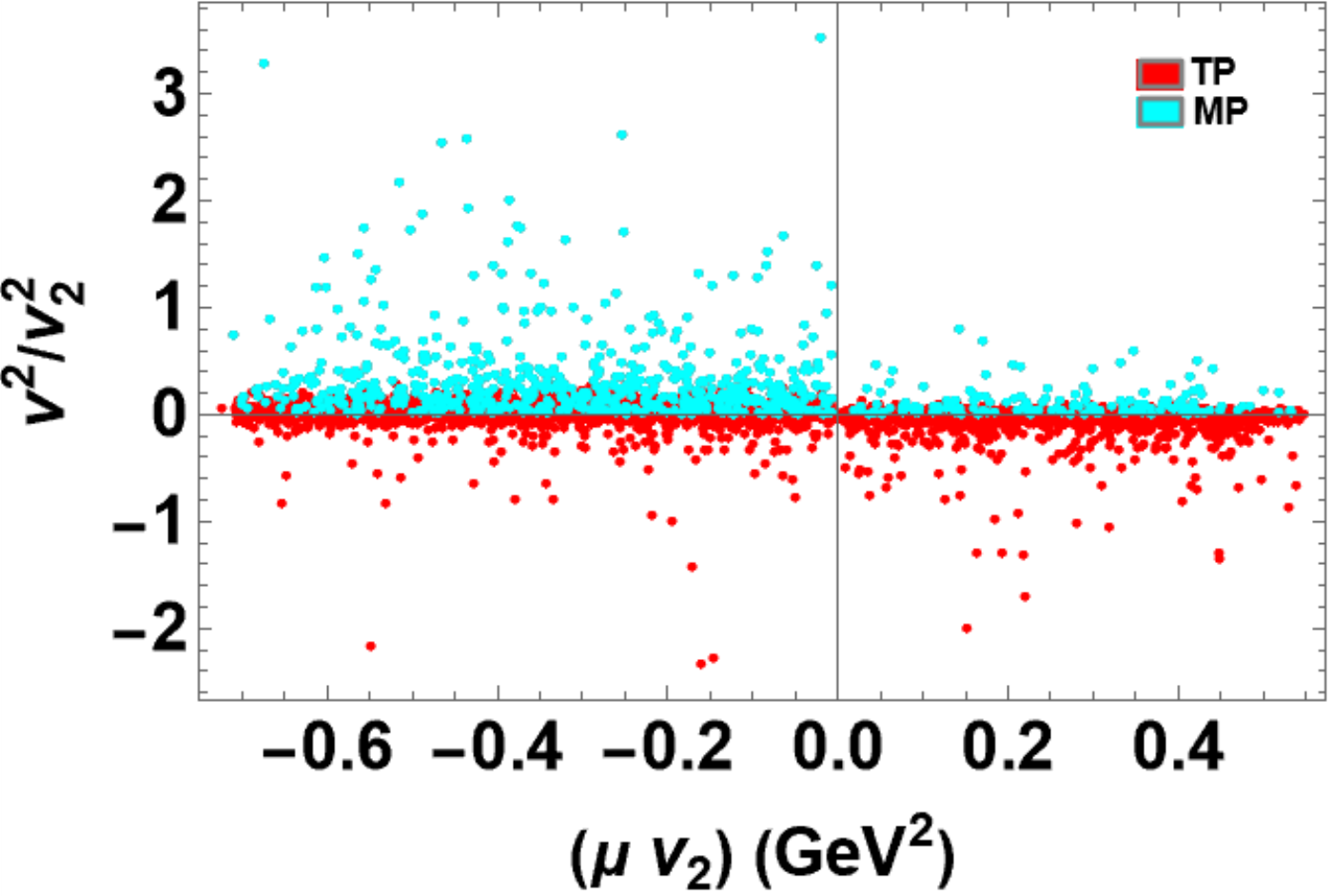}\label{fig:6b}}
   \caption{ Shows the correlation plots between (a) $\lambda_4^{H\Delta}$ vs $v^2/v_2^2$. (b) $v^2/v_2^2$ vs $\mu v_2$.  The red data points signify texture prediction (TP) while the blue ones signify the corrections from model sector(MP).}
\label{fig:pot plots}
\end{figure}

\begin{figure}[H]
  \centering
    \subfigure[]{\includegraphics[width=0.46\textwidth]{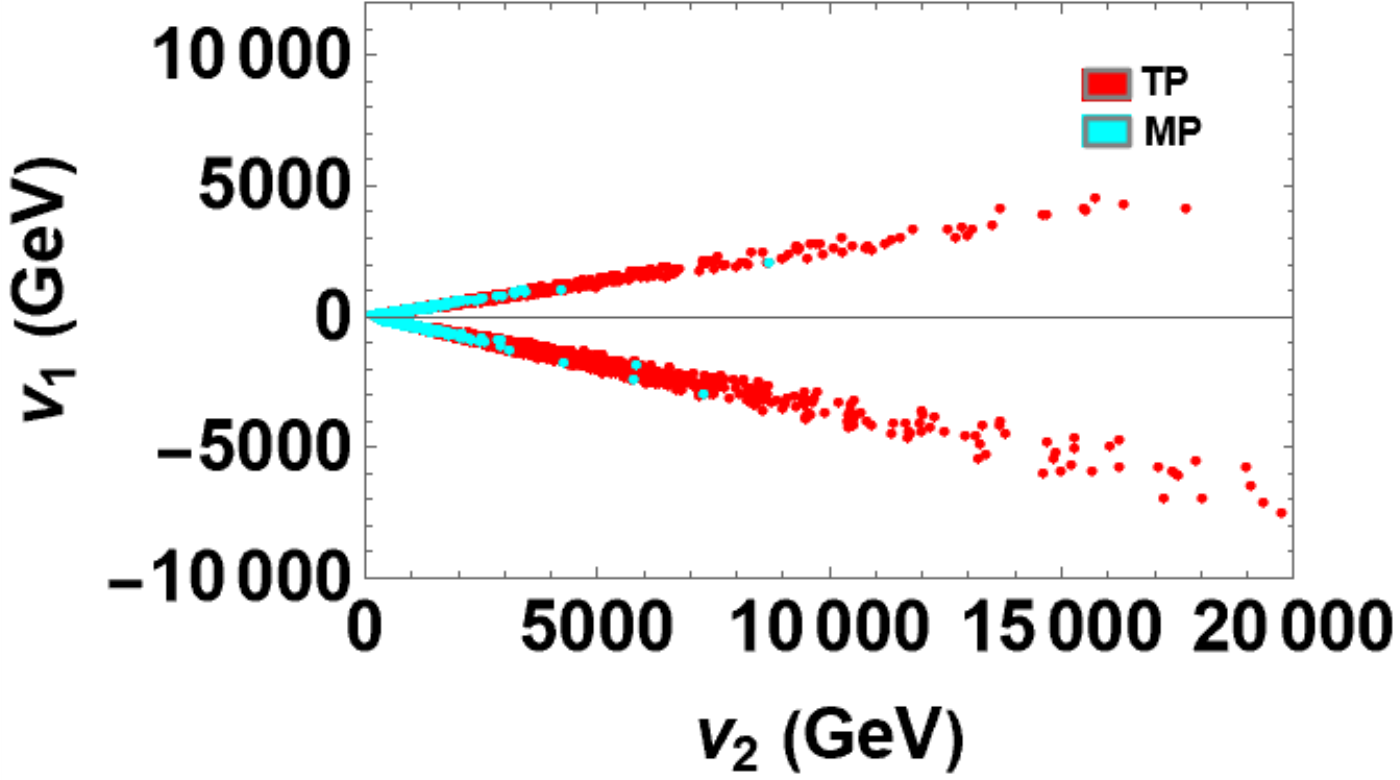}\label{fig:7a}} 
    \subfigure[]{\includegraphics[width=0.45\textwidth]{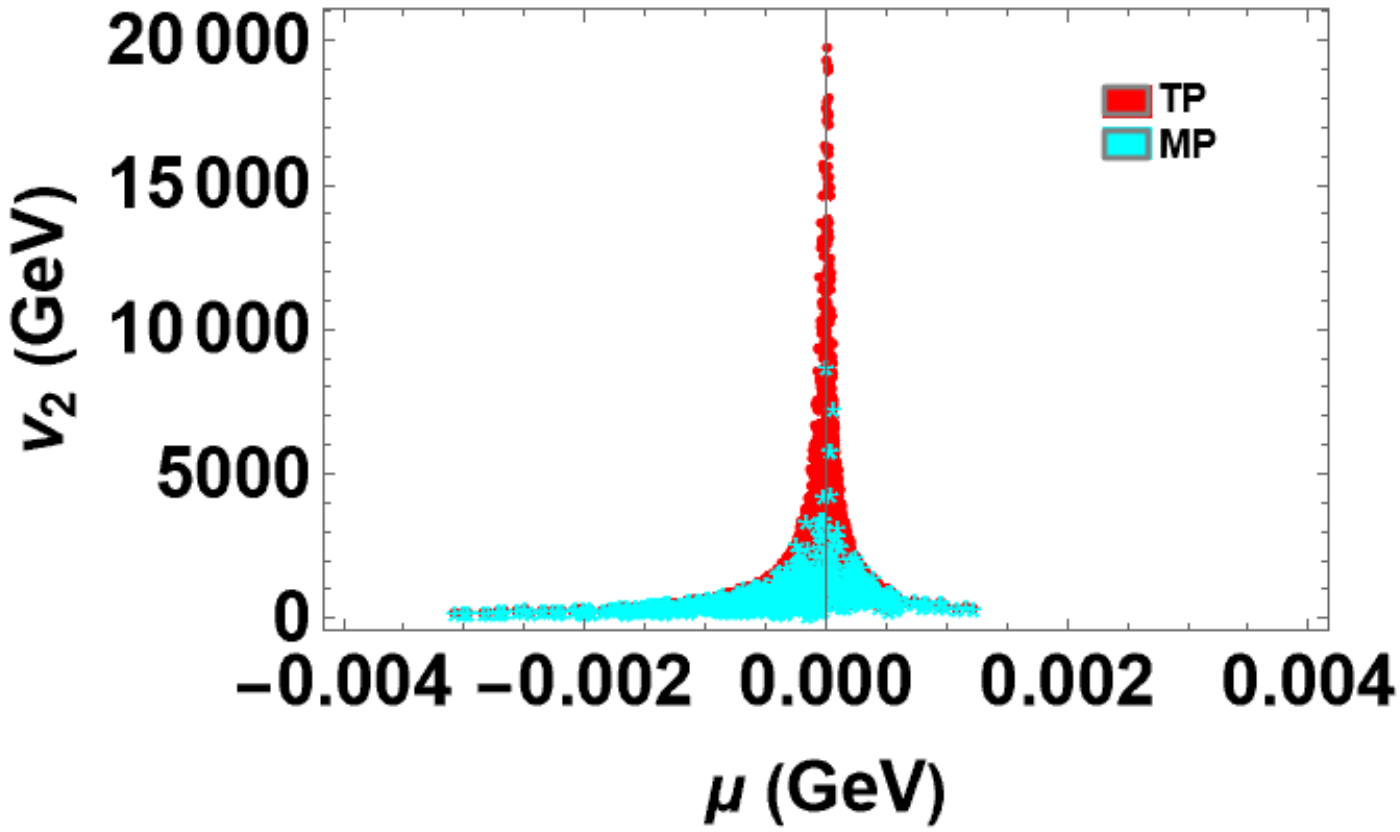}\label{fig:7b}}
   \caption{ Shows the correlation plots between (a) $v_1$ vs $v_2$. (b) $v_2$ vs $\mu$. The red data points signify texture prediction (TP) while the blue ones signify the corrections from model sector(MP).}
\label{fig:v1vsv2}
\end{figure}

\subsection{Refined Parameter Space \label{Relating Yukawa and Scalar Sector}}

We first try to realize the parameter domain of the elements : $\lambda_4^{H\Delta}$, $\mu v_2$ and $(v/v_2)^2$ which depend on the free parameters $\lambda_2^{\Delta}$, $\lambda_3^{\Delta}$, $\lambda_4^{\Delta}$, $\lambda_2^{H \Delta}$, $\lambda_4^{H\Delta}$ and $v_1/v_2$. The parameter $v_1/v_2$ is guided by $r$ and $\theta$. From the phenomenological analysis in section\,\ref{sec2}, we are aware of the parameter space of the two said texture parameters. Moreover, we adhere to the fact that the quartic coupling constants, $\lambda\sim \mathcal{O}(1)$. Motivated by this, we allow the free parameters  $ \lambda_3^{H \Delta}$, $\lambda_2^{\Delta}$, $\lambda_3^{\Delta}$, $ \lambda_4^{\Delta}$ to vary in the range $(0.01-0.99)$. With these inputs, we graphically visualise the parameter space of  $\lambda_4^{H\Delta}$, $\mu v_2$ and $(v/v_2)^2$ in Figs.\,(\ref{fig:pot plots}). We further extend the analysis to gain deeper insight of domain  of $v_1$, $v_2$ and $\mu$ (see Fig.\,(\ref{fig:v1vsv2})).

Our model requires, $(v/v_2)^2 > 0$ and the reality condition demands $\lambda_4^{H\Delta}\lesssim 1$. It is interesting to note that on incorporating these two constraints, the parameter space of $(v/v_2)^2$ is restricted to a narrow region (this is shown in blue in Fig.\,(\ref{fig:6a})). This refinement is not confined only to the said parameter, rather transmits to the parameter space of $v_1$, $v_2$ and $\mu$ (see Fig.\,(\ref{fig:v1vsv2})). 

When we study a texture from a model independent point of view, the parameter space of the observable is quite wide in comparison to the scenario where the former is achieved following a top-down route. This is because the model parameters are subjected to many restrictions. One such scenario is discussed in the earlier paragraph. Thus, the refinement of the model parameters will also affect the predictions of the observables discussed earlier. We revisit the graphical analysis in sections\,\ref{sec3}-\ref{sec4} and highlight the filtered data points in blue (see Figs.\,(\ref{fig:physical parameters})-(\ref{fig:ratio})). In the next section, we shall conclude.

\section{Summary and Discussion \label{sec5}}

The study of neutrino mass matrix texture has been explored exhaustively by the model builders. At the same time, the experiments are gradually achieving more precise predictions, whereas the fundamental issues such as the neutrino mass hierarchy and anomalies related to physical parameters remain unresolved. Hence, there lies the importance of looking into more predictive textures of neutrino mass matrix both from a bottom-up and top-down perspective. Although physics is independent of parametrization, yet a meaningful parametrization may simplify a certain problem and thereby will help to approach a framework in a better way. With this motivation, the present piece of work adapts ``Exponential parametrization'', where the general neutrino mass matrix element is visualised in polar form. We have posited a neutrino mass matrix texture under the roof of the said parametrisation which is quite promising in the light of neutrino oscillation experiments. We predict narrow bound/\,allowed and forbidden regions for the observable parameters like atmospheric angle and three CP violating phases. The texture is perfectly obtained in its exact form ensuring the independence of the parameters, in a discrete symmetry based framework. Moreover in this work, we have emphasized on how accurately the texture is consistent with the model in terms of refined parameter space. It is important to mention that a similar approach encompassing texture analysis and preference for certain parametrization may also assist us in exploring both quark and lepton sectors in the unified frameworks.

In the context of scalar potential and minimization conditions, we would like to state that the strategy is not so straightforward. The present work deals with seven scalar field multiplets as shown in Table (\ref{Field Content of M}), and in reality, there are twenty-nine real scalar fields. Hence, we obtain twenty-nine \emph{optimisation} conditions. To understand whether the chosen vacuum alignment leads to the actual minima, the Hessian matrix:
$H_{ij}=(\frac{\partial^2 V}{\partial \phi_i \partial \phi_j})_0$,
is to be evaluated. The necessary condition for a minimum to occur is to ensure that either the eigenvalues of $H$ are positive (or zero for Goldstone states), or the principal cofactors are positive definite. In the present context, this task is quite challenging, as the Hessian matrix is of dimension twenty-nine. However, we have considered the optimization conditions as the minimization conditions based on the fact that the potential (or the Hessian) involves forty-eight free parameters: one $\mu$, seven $\mu^2$'s, and forty $\lambda$'s. By tuning these parameters appropriately, one may achieve the required conditions. However, this task is beyond the scope of the present work.

\section*{Acknowledgements}
The research work of PC is supported by Innovation in Science Pursuit for Inspired Research (INSPIRE), Department of Science and Technology, Government of India, New Delhi vide grant No. IF190651.

\appendix

\section{Product Rules of $A_4$ \label{appendix a}}

The discrete group $A_4$ is a subgroup of $SU(3)$ and has 12 elements. The group has four irreducible representation out of which three are one dimensional and one is three dimensional. In S-basis, the product of two triplets $(a_1, a_2, a_3)^T$ and $(b_1, b_2, b_3)^T$ gives,

\begin{equation}
3\times3=1+1'+1''+3_S+3_A\nonumber,
\end{equation}

where,

\begin{eqnarray}
(3\times3)_1&=&(a_1 b_1+a_2 b_2+a_3 b_3),\nonumber\\
(3\times3)_{1'}&=&(a_1 b_1+\omega^2 a_2 b_2+\omega a_3 b_3),\nonumber\\
(3\times3)_{1''}&=&(a_1 b_1+\omega a_2 b_2+\omega^2 a_3 b_3),\nonumber\\
(3\times3)_{3_S}&=&\begin{bmatrix}
a_2 b_3+a_3 b_2\\
a_3 b_1+a_1 b_3\\
a_1 b_2+a_2 b_1\\
\end{bmatrix},\nonumber\\
(3\times3)_{3_A}&=&\begin{bmatrix}
a_2 b_3-a_3 b_2\\
a_3 b_1-a_1 b_3\\
a_1 b_2-a_2 b_1\\
\end{bmatrix}\nonumber.
\end{eqnarray}

The trivial singlet can be obtained from the following singlet product rules,

\begin{equation}
1\times1=1,\,\,1'\times1''=1,\,\,1''\times1'=1\nonumber.
\end{equation}

\section{The Scalar Potential \label{appendix b}}

To justify the vacuum alignments of the scalar fields, we construct the $SU(2)_L \times U(1)_Y \times A_4 \times Z_{10} \times Z_7$ invariant scalar potential in the following way,

\begin{eqnarray}
V &=& V(H)+V(\Delta)+V(\chi)+V(\psi)+V(\xi)+V(\eta)+V(\kappa)+V(H,\Delta)\nonumber\\&&+V(H, \chi)+V(H, \psi)+V(H, \xi)+V(H, \eta)+V(H, \kappa)+V(\Delta, \chi)\nonumber\\&&+V(\Delta, \psi)+V(\Delta, \xi)+V(\Delta, \eta)+V(\Delta, \kappa)+V(\chi, \psi)+V(\chi, \xi)\nonumber\\&&+V(\chi, \eta)+V(\chi, \kappa)+V(\psi, \xi)+V(\psi, \eta)+V(\psi, \kappa)+V(\xi, \eta)\nonumber\\&&+V(\xi, \kappa)+V(\eta, \kappa)+\,h.c.\nonumber
\label{potential}
\end{eqnarray}

The explicit forms of the terms appearing in the scalar potential are given by,

\begin{eqnarray}
V(H)&=& \mu^2_H (H^\dagger H)+\lambda^H_1 (H^\dagger H) (H^\dagger H)+\lambda^H_2 (H^\dagger H)_{1'} (H^\dagger H)_{1''}+\nonumber\\&&\lambda^H_3 (H^\dagger  H)_{3_s} (H^\dagger H)_{3_s}+\lambda^H_4 (H^\dagger  H)_{3_s} (H^\dagger H)_{3_a}+\lambda^H_5 (H^\dagger \nonumber\\&& H)_{3_a} (H^\dagger H)_{3_a},\nonumber\\
V(\Delta)&=& \mu^2_\Delta Tr(\Delta^\dagger \Delta)+\lambda^\Delta_1 Tr(\Delta^\dagger \Delta) Tr(\Delta^\dagger\Delta)+\lambda^\Delta_2 Tr(\Delta^\dagger \Delta)_{1'}\nonumber\\&& Tr(\Delta^\dagger \Delta)_{1''}+\lambda^\Delta_3 Tr(\Delta^\dagger \Delta)_{3_s} Tr(\Delta^\dagger \Delta)_{3_s}+\lambda^\Delta_4 Tr(\Delta^\dagger \Delta)_{3_s}\nonumber\\&& Tr(\Delta^\dagger \Delta)_{3_a}+\lambda^\Delta_5 Tr(\Delta^\dagger \Delta)_{3_a} Tr(\Delta^\dagger \Delta)_{3_a},\nonumber\\
V(\chi)&=& \mu^2_\chi (\chi^\dagger \chi)+\lambda^\chi (\chi^\dagger \chi)(\chi^\dagger \chi),\nonumber\\
V(\psi)&=& \mu^2_\psi (\psi^\dagger \psi)+\lambda^\psi (\psi^\dagger \psi)(\psi^\dagger \psi),\nonumber\\
V(\xi)&=& \mu^2_\xi (\xi^\dagger \xi)+\lambda^\xi (\xi^\dagger \xi)(\xi^\dagger \xi),\nonumber\\
V(\eta)&=& \mu^2_\eta (\eta^\dagger \eta)+\lambda^\eta (\eta^\dagger \eta)(\eta^\dagger \eta),\nonumber\\
V(\kappa)&=& \mu^2_\kappa (\kappa^\dagger \kappa)+\lambda^\kappa (\kappa^\dagger \kappa)(\kappa^\dagger \kappa),\nonumber\\
V(H, \Delta )&=& \lambda^{H \Delta}_1 (H^\dagger H)Tr(\Delta^\dagger \Delta)+\lambda^{H \Delta}_2 [(H^\dagger  H)_{1'}Tr(\Delta^\dagger \Delta)_{1''}+\nonumber\\&&(H^\dagger H)_{1''}Tr(\Delta^\dagger \Delta)_{1'}]+\lambda^{H \Delta}_3  (H^\dagger H)_{3_s}Tr(\Delta^\dagger \Delta)_{3_s}+\lambda^{H \Delta}_4 \nonumber\\&&[(H^\dagger H)_{3_s}Tr(\Delta^\dagger \Delta)_{3_a}+(H^\dagger H)_{3_a}Tr(\Delta^\dagger \Delta)_{3_s}]+\lambda^{H \Delta}_5  (H^\dagger \nonumber\\&& H)_{3_a}Tr(\Delta^\dagger \Delta)_{3_a}+\frac{\mu_{H\Delta}}{\Lambda} (H^T i \sigma_2 \Delta^T H \xi),\nonumber\\
V(H, \chi)&=& \lambda^{H \chi}_1 (H^\dagger H)(\chi^\dagger \chi),\nonumber\\
V(H, \psi)&=& \lambda^{H \psi}_1 (H^\dagger H)(\psi^\dagger \psi),\nonumber\\
V(H, \xi)&=& \lambda^{H \xi}_1 (H^\dagger H)(\xi^\dagger \xi),\nonumber\\
V(H, \eta)&=& \lambda^{H \eta}_1 (H^\dagger H)(\eta^\dagger \eta),\nonumber\\
V(H, \kappa)&=& \lambda^{H \kappa}_1 (H^\dagger H)(\kappa^\dagger \kappa),\nonumber\\
V(\Delta, \chi)&=& \lambda^{\Delta \chi}_1 Tr(\Delta^\dagger \Delta)(\chi^\dagger \chi),\nonumber\\
V(\Delta, \psi)&=& \lambda^{\Delta \psi}_1 Tr(\Delta^\dagger \Delta)(\psi^\dagger \psi),\nonumber\\
V(\Delta, \xi)&=& \lambda^{\Delta \xi}_1 Tr(\Delta^\dagger \Delta)(\xi^\dagger \xi)\nonumber,
\end{eqnarray}

\begin{eqnarray}
V(\Delta, \eta)&=& \lambda^{\Delta \eta}_1 Tr(\Delta^\dagger \Delta)(\eta^\dagger \eta),\nonumber\\
V(\Delta, \kappa)&=& \lambda^{\Delta \kappa}_1 Tr(\Delta^\dagger \Delta)(\kappa^\dagger \kappa),\nonumber\\
V(\chi, \psi)&=& \lambda^{\chi \psi}_1 (\chi^\dagger \chi)(\psi^\dagger \psi),\nonumber\\
V(\chi, \xi)&=& \lambda^{\chi \xi}_1 (\chi^\dagger \chi)(\xi^\dagger \xi),\nonumber\\
V(\chi, \eta)&=& \lambda^{\chi \eta}_1 (\chi^\dagger \chi)(\eta^\dagger \eta),\nonumber\\
V(\chi, \kappa)&=& \lambda^{\chi \kappa}_1 (\chi^\dagger \chi)(\kappa^\dagger \kappa),\nonumber\\
V(\psi, \xi)&=& \lambda^{\psi \xi}_1 (\psi^\dagger \psi)(\xi^\dagger \xi),\nonumber\\
V(\psi, \eta)&=& \lambda^{\psi \eta}_1 (\psi^\dagger \psi)(\eta^\dagger \eta),\nonumber\\
V(\psi, \kappa)&=& \lambda^{\psi \kappa}_1 (\psi^\dagger \psi)(\kappa^\dagger \kappa),\nonumber\\
V(\xi, \eta)&=& \lambda^{\xi \eta}_1 (\xi^\dagger \xi)(\eta^\dagger \eta),\nonumber\\
V(\xi, \kappa)&=& \lambda^{\xi \kappa}_1 (\xi^\dagger \xi)(\kappa^\dagger \kappa),\nonumber\\
V(\eta, \kappa)&=& \lambda^{\eta \kappa}_1 (\eta^\dagger \eta)(\kappa^\dagger \kappa)\nonumber.
\end{eqnarray}

For simplicity, we redefine $\mu=\frac{1}{\Lambda} \mu_{H\Delta} \xi$. Without loss of generality, the following equations are valid from the minimisation conditions of the scalar potential for the chosen vacuum alignments: $\langle H \rangle_{0}=v_{H}(1,1,1)^{T}$, $\langle \Delta \rangle_{0}=v_{\Delta}(v_1,v_2,0)^{T}$, $\langle\chi\rangle_{0}=v_{\chi}$, $\langle\psi\rangle_{0}=v_{\psi}$, $\langle\xi\rangle_{0}=v_{\xi}$, $\langle\eta\rangle_{0}=v_{\eta}$ and $\langle\kappa\rangle_{0}=v_{\kappa}$,

\begin{eqnarray}
 v_H (-\mu_H^2+ v_H^2 (6 \lambda_1^H+8 \lambda_3^H)+v_1^2 (\lambda_1^{H\Delta}+2\lambda_2^{H\Delta})+2 v_1 v_2 (\lambda_3^{H\Delta}\nonumber\\-\lambda_4^{H\Delta})+v_2 (2 \mu +\lambda_1^{H\Delta} v_2-\lambda_2^{H\Delta} v_2)+\lambda_1^{H\eta}v_\eta^2+\lambda_1^{H\kappa} v_\kappa^2+\lambda_1^{H\xi}\nonumber\\ v_\xi^2+\lambda_1^{H\chi} v_\chi^2+\lambda_1^{H\psi} v_\psi^2)=0,\label{b30}
\end{eqnarray}
\begin{eqnarray}
 v_H (-\mu_H^2 + v_H^2 (6 \lambda_1^{H} + 8 \lambda_3^{H}) 
+ v_1^2 (\lambda_1^{H\Delta} - \lambda_2^{H\Delta}) 
+ 2 v_1 v_2 (\lambda_3^{H\Delta}\nonumber\\ + \lambda_4^{H\Delta}) 
+ v_2^2 (\lambda_1^{H\Delta} + 2 \lambda_2^{H\Delta}) 
+ \lambda_1^{H\eta} v_{\eta}^2 
+ \lambda_1^{H\kappa} v_{\kappa}^2 
+ \lambda_1^{H\xi} v_{\xi}^2 
+ \nonumber\\\lambda_1^{H\chi} v_{\chi}^2 
+ \lambda_1^{H\psi} v_{\psi}^2)=0
\end{eqnarray}
\begin{eqnarray}
v_H( -\mu_H^2 + v_H^2 (6 \lambda_1^H + 8 \lambda_3^H) + (\lambda_1^{H\Delta} - \lambda_2^{H\Delta}) (v_1^2 + v_2^2) + 2 \mu v_1 \nonumber\\ + \lambda_1^{H\eta} v_\eta^2 + \lambda_1^{H\kappa} \, v_\kappa^2 + \lambda_1^{H\xi} \, v_\xi^2 + \lambda_1^{H\chi} \, v_\chi^2 + \lambda_1^{H\psi} v_\psi^2)=0,
\end{eqnarray} 
\begin{eqnarray}
v_H^2 ( 3 \lambda_1^{H\Delta} v_1 + 2 (\mu + \lambda_3^{H\Delta} v_2 - \lambda_4^{H\Delta} v_2)) 
+ v_1 ( -\mu_{\Delta}^2 + 2 (\lambda_1^{\Delta}+\lambda_2^{\Delta}) v_1^2 
\nonumber\\+ 2 v_2^2 (\lambda_1^{\Delta}-\lambda_2^{\Delta} + 2 \lambda_3^{\Delta} - \lambda_4^{\Delta}) 
+ \lambda_1^{\Delta \eta} v_{\eta}^2 
+ \lambda_1^{\Delta \kappa} v_{\kappa}^2 
+ \lambda_1^{\Delta \xi} v_{\xi}^2 
+ \lambda_1^{\Delta \chi} \nonumber\\v_{\chi}^2 
+ \lambda_1^{\Delta \psi} v_{\psi}^2)=0,\label{dvd1}
\end{eqnarray} 
\begin{eqnarray}
v_H^2 (2 (\mu + v_1 (\lambda_3^{H\Delta} + \lambda_4^{H\Delta})) + 3 \lambda_1^{H\Delta} v_2 )  
+ v_2 (-\mu_{\Delta}^2 + 2 v_1^2 (\lambda_1^{\Delta}-\lambda_2^{\Delta}+\nonumber\\ 2 \lambda_3^{\Delta} + \lambda_4^{\Delta})  
+ 2 (\lambda_1^{\Delta}+\lambda_2^{\Delta}) v_2^2 + \lambda_1^{\Delta \eta} v_{\eta}^2  
+ \lambda_1^{\Delta \kappa} v_{\kappa}^2 + \lambda_1^{\Delta \xi} v_{\xi}^2  
+ \lambda_1^{\Delta \chi} \nonumber\\v_{\chi}^2 + \lambda_1^{\Delta \psi} v_{\psi}^2)=0,
\end{eqnarray}
\begin{eqnarray}
2 v_H^2 (\mu + v_1 (\lambda_3^{H\Delta} - \lambda_4^{H\Delta}) + v_2 (\lambda_3^{H\Delta} + \lambda_4^{H\Delta}))=0,\label{dvd3}
\end{eqnarray}

\begin{eqnarray}
 v_{\chi} (-\mu_{\chi}^2 + 3 \lambda_1^{H\chi} v_H^2 + \lambda_1^{\Delta\chi} (v_1^2 + v_2^2) + \lambda_1^{\chi\eta} v_{\eta}^2 + \lambda_1^{\chi\kappa} v_{\kappa}^2 + \lambda_1^{\chi\xi} v_{\xi}^2  \nonumber\\ +2 \lambda_1^{\chi} v_{\chi}^2 + \lambda_1^{\chi\psi} v_{\psi}^2 )=0,
\end{eqnarray}
\begin{eqnarray}
v_{\psi} ( -\mu_{\psi}^2 + 3 \lambda_1^{H\psi} v_H^2 + \lambda_1^{\Delta\psi} (v_1^2 + v_2^2) 
+ \lambda_1{\psi\eta} v_{\eta}^2 + \lambda_1^{\psi\kappa} v_{\kappa}^2+ \lambda_1^{\psi\xi} v_{\xi}^2\nonumber\\ + \lambda_1^{\chi\psi} v_{\chi}^2 + 2 \lambda_{\psi} v_{\psi}^2 )=0,
\end{eqnarray}
\begin{eqnarray}
v_{\xi} (-\mu_{\xi}^2 + 3 \lambda_1^{H\xi} v_H^2 + \lambda_1^{\Delta\xi} (v_1^2 + v_2^2) + \lambda_1^{\xi\eta} v_{\eta}^2 + \lambda_1^{\xi\kappa} v_{\kappa}^2 + 2\lambda_{\xi} v_{\xi}^2 +\nonumber\\ \lambda_1^{\chi\xi} v_{\chi}^2 + \lambda_1^{\psi\xi} v_{\psi}^2 )=0,
\end{eqnarray}
\begin{eqnarray}
v_{\eta } (-\mu _{\eta }^2+2 \lambda _1^{\Delta \eta } (v_1^2 + v_2^2)+2 \lambda ^{\eta } v_{\eta }^2+\lambda _1^{\eta \kappa } v_{\kappa }^2+\lambda _1^{\xi \eta } v_{\xi }^2+\lambda _1^{\chi \eta } v_{\chi }^2+\nonumber\\ \lambda _1^{\psi \eta } v_{\psi }^2+3 \lambda _1^{H \eta } v_{H }^2)=0,
\end{eqnarray}
\begin{eqnarray}
v_{\kappa } (-\mu _{\kappa }^2+2 \lambda _1^{\Delta \kappa } (v_1^2 + v_2^2)+\lambda _1^{\eta \kappa } v_{\eta }^2+2 \lambda ^{\kappa } v_{\kappa }^2+\lambda _1^{\xi \kappa } v_{\xi }^2+\lambda _1^{\chi \kappa } v_{\chi }^2+\nonumber\\\lambda _1^{\psi \kappa } v_{\psi }^2+3 \lambda _1^{H \kappa } v_{H }^2)=0\label{b40}.
\end{eqnarray}

\vspace{1 cm}
 
 \bibliographystyle{JHEP}
\bibliography{ref.bib}

\end{document}